\documentclass[10pt]{article}
\usepackage{amsmath}
\usepackage{amssymb}
\usepackage{times}
\usepackage{graphicx}
\usepackage{color}
\usepackage{multirow}
\usepackage{url}
\usepackage{hyperref}
\usepackage{listings}
\usepackage{rotating}
\usepackage{bbm}
\usepackage{latexsym}

\usepackage{xr}
\usepackage{hyperref}

\newcommand{\captionfonts}{\normalsize}

\makeatletter  
\long\def\@makecaption#1#2{%
  \vskip\abovecaptionskip
  \sbox\@tempboxa{{\captionfonts #1: #2}}%
  \ifdim \wd\@tempboxa >\hsize
    {\captionfonts #1: #2\par}
  \else
    \hbox to\hsize{\hfil\box\@tempboxa\hfil}%
  \fi
  \vskip\belowcaptionskip}
\makeatother   
\definecolor{codegreen}{rgb}{0,0.6,0}
\definecolor{codegray}{rgb}{0.5,0.5,0.5}
\definecolor{codepurple}{rgb}{0.58,0,0.82}
\definecolor{backcolour}{rgb}{0.95,0.95,0.92}

\lstdefinestyle{mystyle}{
    backgroundcolor=\color{backcolour},   
    commentstyle=\color{codegreen},
    keywordstyle=\color{magenta},
    numberstyle=\tiny\color{codegray},
    stringstyle=\color{codepurple},
    basicstyle=\ttfamily\footnotesize,
    breakatwhitespace=false,         
    breaklines=true,                 
    captionpos=b,                    
    keepspaces=true,                 
    numbers=left,                    
    numbersep=5pt,                  
    showspaces=false,                
    showstringspaces=false,
    showtabs=false,                  
    tabsize=2
}

\newcommand{\cv}[1]{{\bf #1}}
\newcommand{\cvsym}[1]{{\boldsymbol{#1}}}
\newcommand{\transp}{^{\mathrm{T}}}
\newcommand{\mat}[1]{{\bf #1}}
\newcommand{\setR}{{\mathbb{R}}}

\newcommand{\tightitems}{\itemsep0pt\topsep0pt}

\begin{document}
\hspace{13.9cm}1

\ \vspace{20mm}\\

{\LARGE IVISIT: An Interactive Visual Simulation Tool for system simulation, visualization, optimization, and parameter management}

\ \\
{\bf \large Andreas Knoblauch$^{\displaystyle 1}$}\\
{$^{\displaystyle 1}$Albstadt-Sigmaringen University, KEIM Institute, Poststrasse 6, 72458 Albstadt-Ebingen, Germany, knoblauch@hs-albsig.de}\\
%{$^{\displaystyle 2}$Your second affiliation.}\\
%

%\ \\[-2mm]

\thispagestyle{empty}
\markboth{}{NC instructions}
\ \vspace{-0mm}\\
%
%Abstract
\begin{center} {\bf Abstract} \end{center}
IVISIT is a generic interactive visual simulation tool that is based on Python/Numpy and
can be used for system simulation, parameter optimization, parameter management, and visualization of system dynamics as required, for example,
for developing neural network simulations, machine learning applications, or computer vision systems. It provides classes for rapid prototyping
of applications and visualization and manipulation of system properties using interactive GUI elements like sliders, images, textboxes, option lists, checkboxes
and buttons based on Tkinter and Matplotlib. Parameters and simulation configurations can be stored and managed based on SQLite database functions.
This technical report describes the main architecture and functions of IVISIT, and provides easy examples
how to rapidly implement interactive applications and manage parameter settings.
%%%%%%%%%%%

\tableofcontents

\vspace*{0.4cm}\noindent All {\bf Python sources} and parameter files for the {\bf IVISIT examples} in section~3
can be downloaded from the link ``TeX Source'' on \url{https://www.arxiv.org/abs/2408.03341}.

\newpage

% *******************************************************************************************
% *******************************************************************************************
\section{Introduction}  \label{sec:introduction}
% *******************************************************************************************
% *******************************************************************************************
When simulating neural networks of the brain or optimizing artificial neural networks for machine learning and Artificial Intelligence (AI) there is a strong demand
for an intuitive understanding of the properties and dynamics of the system under investigation. For this, a visualization of system states is very helpful, and 
an interaction with the system by online manipulation of system parameters often provides additional insights. There exist already a number of software tools to support such functions. For example, Jupyter Notebooks are often used for integrating python code, data visualization and documentation, for example, for machine learning and statistics applications
\cite{Kluyver:jupyter:2016,Rossum/Drake:Python3:2009,Silaparasetty:JupyterMachineLearning:2020}.
Also Matlab has introduced LiveScripts that support a similar functionality \cite{MATLAB:2022}. IVISIT provides similar functions
based on Python/Numpy \cite{Rossum/Drake:Python3:2009,PythonOrg:2023,Harris_etal_Numpy:2020},
and integrates interactive visual elements based on the Tkinter GUI \cite{Tkinter:2023}
with functions for managing parameter settings and simulation configurations based on the SQLite database framework \cite{Hipp:SQLite:2020}.
The general application scenario for IVISIT
is to investigate a system (for example, neural networks, computer vision application, machine learning models) by interactively manipulating and
optimizing system parameters using visual feedback, saving parameter settings, trying and saving other parameter configurations, and evaluating the best fitting parameters.
Examples for IVISIT applications are simulations of biologically realistic neural networks, e.g., using IVISIT as a front end for the C/C++ based
Felix++ simulation tool \cite{Knoblauch:2003_b,Wennekers/Garagnani/Pulvermuller:2006,Wennekers:1999}, using IVISIT to create and optimize machine learning applications, labeling tools for machine learning, or applications for conducting psychophysical
experiments. IVISIT applications have also been used successfully for teaching and practical exercises, for example, in lectures on computer vision, machine learning, and control.

This technical report describes the main architecture and functions of IVISIT, and provides easy-to-understand examples
how to rapidly implement interactive applications and manage parameter settings. Section~\ref{sec:architecture} gives an overview on the architecture of IVISIT, including
brief descriptions of the Python modules and their interdependencies. For a {\bf quick start}, section~\ref{sec:examples} gives several examples how to implement
IVISIT simulation applications. In particular, section~\ref{sec:example1_HelloWorld} gives a ``hello-world'' example, including hints for {\bf installation}
and a {\bf best-practice} overview of the IVISIT application environment and functions,
including parsing of GUI elements, running and stopping a simulation, and loading and storing simulation parameters and GUI settings.
After that, sections~\ref{sec:example2_LIFNeuron_simplescope}-\ref{sec:example6_InteractiveTrainingKerasMNIST} exemplarily describe more {\bf advanced features}:
Section~\ref{sec:example2_LIFNeuron_simplescope} shows a simulation of Leaky-Integrate-And-Fire (LIF) neuron model to explain how to use a \underline{SimpleScope} to display
time-recordings of signals like the dendritic potential and spikes of neurons in an oscilloscope-like device.
Section~\ref{sec:example3_LIFNeuron_matplotlib} shows a similar simulation of a LIF-neuron, but plotting
neural activities by connecting to the \underline{Matplolib} library \cite{Hunter:Matplotlib:2007}.
Section~\ref{sec:example4_InteractiveMatplotlibGUI} shows an example how to use the \underline{Tkinter event handling framework} to create an interactive
GUI element to create, move, and delete data points in a Matplotlib figure. Section~\ref{sec:example5_InteractiveGUI_IVISIT_eventautomatons} presents a similar IVISIT application
for classification of data points, but where data points are interactively created, moved, and deleted using an \underline{IVISIT event handling automaton} (ClickDragEventAutomaton),
which is much simpler to implement than using the Tkinter event handling framework. As the last example, section~\ref{sec:example6_InteractiveTrainingKerasMNIST} presents
an IVISIT application for interactive training of a \underline{Convolutional Neural Network (CNN)} based on the \underline{Tensorflow/Keras} framework \cite{AbadiTensorflow:2016,CholletKeras:2015}
and using the \underline{MNIST data set}
of handwritten digits \cite{LeCun/Bottou/Bengio/Haffner:MNIST:1998,Deng:MNIST:2012}.
Here the images of the MNIST data set can be visualized interactively, and new images can be easily created and tested by drawing into an image widget,
copying or overlaying MNIST images, and doing transformations like shifting, rotating, and adding noise. 
Finally, section~\ref{sec:conclusions} gives some brief conclusions. Appendix~\ref{app:UML_class_diagrams}
shows {\bf UML class diagrams} and the definitions of the {\bf SQL tables}. And appendix~\ref{app:InCodeCommandsGUIWidgets}
holds an overview of {\bf in-code Python commands to create IVISIT GUI widgets}.

% *******************************************************************************************
% *******************************************************************************************
\section{Overview architecture}  \label{sec:architecture}
\begin{figure}[ht]
%\nopagenumber
%\renewcommand{\baselinestretch}{1.0}
\hfill
\begin{center}
\includegraphics[width=\linewidth]{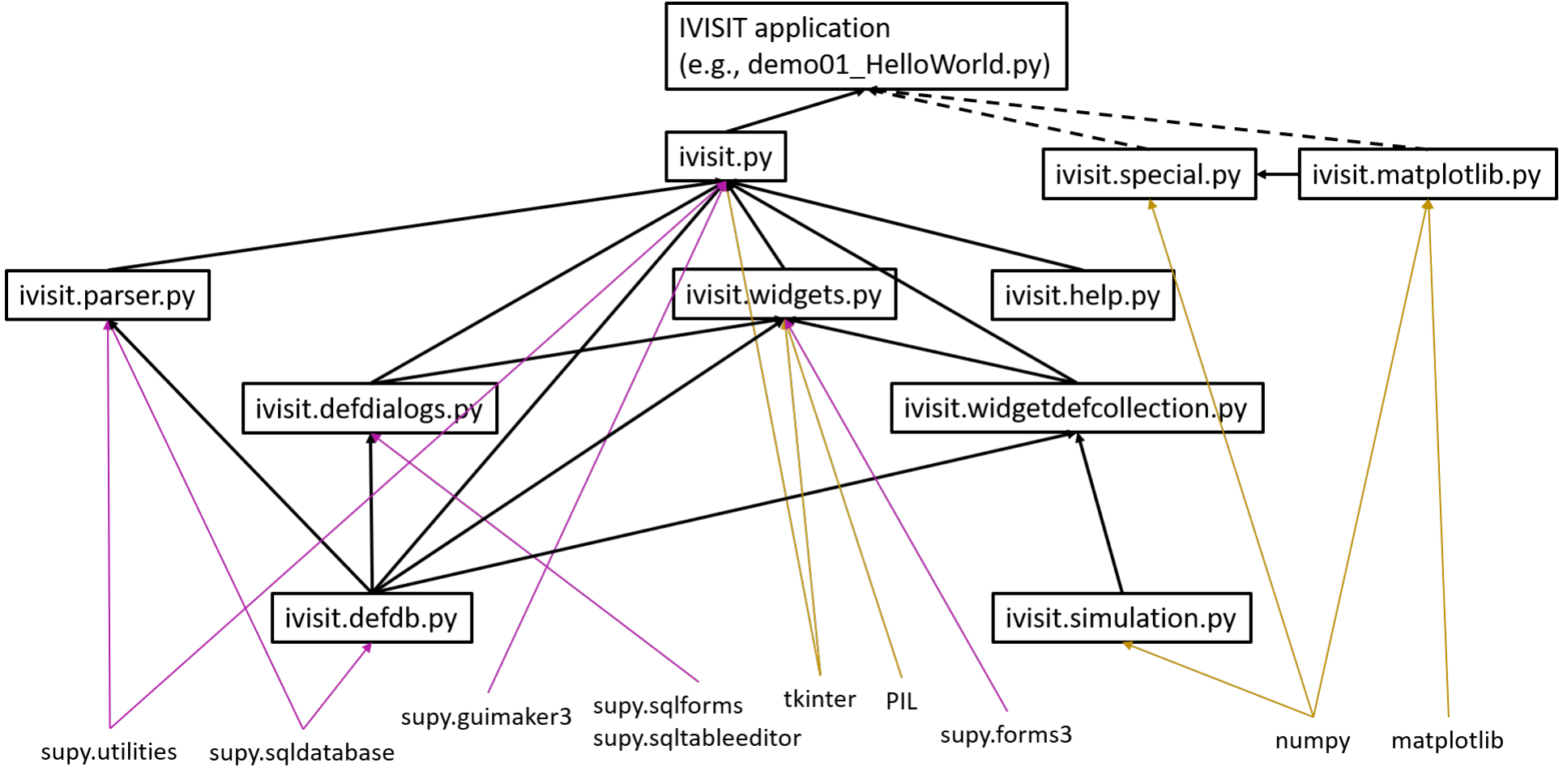}    
\end{center}
\caption{\label{fig:IVISIT_architecture}
  Architecture of IVISIT. The graph shows all IVISIT Python modules (black boxes) and their mutual dependencies, as well as the dependencies to
  external (non-standard) Python libraries (like Numpy, Matplotlib, etc.).
} 
\end{figure}

% *******************************************************************************************
% *******************************************************************************************
The general architecture in terms of Python modules is displayed in Fig.~\ref{fig:IVISIT_architecture}. The main IVISIT modules are displayed as rectangular boxes. Auxiliary Python modules that are not part of a Python standard installation are displayed in smaller type (without boxes). Dependencies between modules are illustrated by arrows.   
To write a simple IVISIT application (e.g., {\tt demo01\_HelloWorld.py}, see section~\ref{sec:example1_HelloWorld})
it is only necessary to import the module {\tt ivisit.py}. Optionally, the modules {\tt ivisit.special.py} and
{\tt ivisit.matplotlib.py} may be included to create additional visual elements. The following briefly explains the role of each module:

% Figure fig1_IVISIT_architecture
% created with PowerPoint on HS-Laptop
% --> see: Figur 1
%     in \Users\knoblauch\Desktop\Aktuelles_Draft\KEIM-Projekte\KICAD-CMCENgineers-hermle\Berichte\ivisit_techreport_materials\TechreportIVISIT_Figuren.pptx 
%     Kopie davon liegt auf: ~/hs-albsig/02-research/KICAD/ivisit_techreport/figures/ivisit_techreport_materials
% exportiert als png von PowerPoint; umgewandelt mit img2pdf in pdf

\begin{itemize}\tightitems
\item {\bf IVISIT application, e.g. demo01\_HelloWorld.py}: The purpose of the application module is to create a GUI application to simulate some system and visualize
      and/or optimize it using the IVISIT GUI elements.  Any application module must import at least {\tt ivisit.py} to overwrite the base classes
      {\tt IVisit\_Parameters}, {\tt IVisit\_Data}, and {\tt IVisit\_Simulation} from the module {\tt ivisit.simulation.py}:
      \begin{itemize}\tightitems
      \item {\tt IVisit\_Parameters}: Defines an interface from IVISIT GUI elements to Python variables. All Python variables defined in a class derived
            from {\tt IVisit\_Parameters} are visible for IVISIT an can be manipulated by GUI elements like sliders, text input fields, option list,
            radiobuttons, or checkboxes.
      \item {\tt IVisit\_Data}: Defines an interface from Python variables to IVISIT GUI elements. All Python variables defined in a class derived
            from {\tt IVisit\_Data} are visible for IVISIT an can be displayed by GUI elements like images or text output fields. 
      \item {\tt IVisit\_Simulation}: Defines a class for simulation objects. The constructor of the class must create and initialize the system to be simulated.
            Besides, it should have an {\tt init()} function to initialize (or re-initialize) all systems states and a  {\tt step()} function to
            update all system states during one simulation step. Optionally, it may contain a {\tt bind()} function to link GUI events to the simulation (e.g., to bind
            mouse clicks on image pixels to some functions in the application that will manipulate these pixels).
      \end{itemize}
      When starting the IVISIT application (e.g., {\tt demo01\_HelloWorld.py}, see section~\ref{sec:example1_HelloWorld}) this will create the simulation object.
      Clicking on the {\tt Init} button will evoke a call to the simulations {\tt init()} functions. Similarly, clicking the {\tt Step} button will
      call the {\tt step()} function. Pressing the {\tt Run} button will automatically call the {\tt step()} function repeatedly. For more details
      see section~\ref{sec:example1_HelloWorld}.
\item {\bf ivisit.py}: This is the main IVISIT module that defines the general look-and-feel of an IVISIT application. The Class {\tt IVisit} defines all menu items and
      implements the respective functions. It also defines all action buttons (on the bottom of the application window) and file dialogs to load and save parameters from/to
      the parameter database files ({\tt $\ast$.db}). For example, for the action buttons {\tt Init}, {\tt Step}, {\tt Run}, {\tt Stop}, and {\tt Continue} see the functions
      {\tt onSim\_main\_init()}, {\tt onSim\_init()}, {\tt onSim\_step()}{\tt onSim\_run()}, {\tt onSim\_stop()}, and {\tt onSim\_cont()}.
      The class {\tt IVisit\-Main} combines this with a GUI application window. In fact, any IVISIT applications will first define a simulation class (e.g., {\tt Sim}) derived
      from {\tt IVisit\_Simulation} as explained in the previous item, and the start the GUI application using {\tt ivisit.IVisit\_main(sim=Sim())}.
      See section~\ref{sec:example1_HelloWorld} for a simple example.
\item {\bf ivisit.special.py}:  \label{txt:item:ivisist_special}
      Contains additional convenience functions and classes for displaying simulation data with the GUI elements of IVISIT and doing mouse event handling using the Tkinter event handling framework.
      For example, class {\tt SimpleScope}
      provides an ``oscilloscope'' to display time-dependent data on screen with IVISIT image objects. For this, calling {\tt SimpleScope. set\_data(x,y)} in
      a simulation step will display data value {\tt y} at current simulation time {\tt x} on the oscilloscope screen. For a concrete example,
      see section~\ref{sec:example2_LIFNeuron_simplescope}. As another example, class {\tt ClickDragEventAutomaton} implements an event handling automaton for easy processing
      of Tkinter mouse click and move events. For a concrete example, see section~\ref{sec:example5_InteractiveGUI_IVISIT_eventautomatons} where data points
      of different classes are created interactively and used to train a classifier. 
\item {\bf ivisit.matplotlib.py}:  \label{txt:item:ivisist_matplotlib}
      Contains interface functions with the Matplotlib library \cite{Hunter:Matplotlib:2007}. With this, you can create high-quality Matplotlib figures and display them
      in interactive IVISIT applications. Call {\tt getMatplotlibFigure()} to get a Matplotlib figure to be drawn within the simulation {\tt step()},
      and call {\tt getMatplotlibImage()} to extract an image matrix from the Matplotlib figure canvas that can be displayed in IVISIT GUI widgets like {\tt @IVISIT:IMAGE}.
      The module provides also the functions {\tt getDataPos\_from\_PixelPos()} and {\tt getDataPos\_from\_ canvas()} to transform between canvas pixel positions and
      Matplotlib coordinates. This is useful, for example, when implementing interactive GUI elements where you need to evaluate click events on Matplotlib figures.
      For an elementary Matplotlib example see section~\ref{sec:example3_LIFNeuron_matplotlib}. For an advanced example involving interactive Matplotlib GUI elements and
      coordinate transforms see section~\ref{sec:example4_InteractiveMatplotlibGUI}. 
\item {\bf ivisit.help.py}: Defines the text for the help menu in IVISIT applications when clicking menu item {\tt Help/Help\_on\_IVISIT}.
      The help text for {\tt Help/About\_IVISIT} is defined in the main module {\tt ivisit.py}.
      You can also provide additional help text for your IVISIT application by starting it using
      {\tt ivisit.IVisit\_main(sim=Sim(),str\_app\_name='appname',str\_app \_help='helptext',str\_app\_about='abouttext')}
      (see also previous item on {\tt ivisit .py}).
      Here {\tt appname} is the name of your app and will be displayed in the windows title and the help menu,
      {\tt helptext} will be displayed when clicking menu item {\tt Help/Help\_on\_appname},
      and {\tt abouttext} will be displayed when clicking on menu item {\tt Help/About\_appname}. For a simple example
      see section~\ref{sec:example1_HelloWorld}.
\item {\bf ivisit.widgets.py}:\label{item:modules_ivisit_widgets} Defines classes for GUI widgets and corresponding container classes.
      All widgets are derived from base
      class {\tt IVisitWidget}. There are three types of widgets (see class diagrams of Fig.~\ref{fig:design_gensim}
      in appendix~\ref{app:UML_class_diagrams}).
      \begin{itemize}\tightitems
      \item {\bf Parameter Widgets:} Cast Python parameter variables of IVISIT applications as Tkinter GUI elements to enable interactive
            simulation and optimization. Corresponding subclasses are {\tt IVisitSliderWidget}, {\tt IVisitDictSliderWidget},
            {\tt IVisitListSelection\-Widget}, {\tt IVisitTextInputWidget}, {\tt IVisitCheckboxWidget}, {\tt IVisitRadio\-buttonWidget},
            and {\tt IVisitButton Widget}.
      \item {\bf Data Widgets:} Cast Python data variables of IVISIT appliations as Tkinter GUI elements for displaying and monitoring of
            system states or evaluations of the simulations. Corresponding subclasses are {\tt DataImage} and {\tt DataTxtField}.
      \item {\bf Comment Widgets} enable displaying additional information and comments as TKinter GUI elements. Currently, the only subclass
            is {\tt IVisitTextCommentWidget}.
      \end{itemize}
      In addition, the module contains classes {\tt IVisitWidgetFactory} and {\tt IVisitRawDisplay Frame}
      for creating and displaying widgets:
      \begin{itemize}
      \item Calling {\tt IVisitWidgetFactory.create\_IVisitWidget(parent,widgetdef)} creates
            a single {\tt IVisitWidget} as defined by {\tt widgetdef} (see {\tt ivisit.widgetdef\-collection.py}) and adds it to {\tt parent}
            which is typically a TKinter Frame/container object holding and displaying the IVISIT widgets. 
      \item Calling the constructor {\tt IVisitRawDisplayFrame(parent, wdefcoll)} defines a TKinter Frame that creates and stores
            all the {\tt IVisitWidget} objects defined by the ``widget definition collection'' {\it wdefcoll}
            (see module {\tt ivisit.widgetdefcollection.py}).
            The latter contains all definitions of widgets from the SQL tables
            {\tt tb\_parameterwidget}, {\tt tb\_datawidget}, and {\tt tb\_commentwidget} associated with the currently active simulation
            defined in {\tt tb\_simulation}
            (see Fig.~\ref{fig:IVISIT_SQL_tables} in appendix~\ref{app:UML_class_diagrams}; see module {\tt ivisit.defdb. py}). 
      \end{itemize}
      {\tt IVisitWidget} objects can be defined directly via the SQL tables, or more conveniently,
      by additional code elements directly in the Python code of the IVISIT application
      that can be parsed by {\tt ivisit.parser.py} (see examples in section~\ref{sec:examples}).
      Appendix~\ref{app:InCodeCommandsGUIWidgets} gives an overview of the in-code commands to create {\tt IVisitWidget} objects.
\item {\bf ivisit.widgetdefcollection.py}: Provides classes defining the IVISIT widgets created by {\tt ivisit. widgets.py} and interfacing with the
      information stored persistently in the SQL tables (see {\tt ivisit.defdb.py} and Fig.~\ref{fig:IVISIT_SQL_tables} in appendix~\ref{app:UML_class_diagrams}).
      For each widget type there is a corresponding class holding its attributes taken from the corresponding SQL table:
      \begin{itemize}\tightitems
      \item {\tt ParameterWidgetDef}: For each {\bf Parameter Widget} from the GUI of an IVISIT application, a corresponding object from class {\tt ParameterWidgetDef} contains the
            defining attributes from SQL table {\tt tb\_parameterwidget} (Fig.~\ref{fig:IVISIT_SQL_tables}) to allow the creation of the GUI widget (see {\tt ivisit.widgets}).
      \item {\tt DataWidgetDef}: For each {\bf Data Widget} from the GUI of an IVISIT application, a corresponding object from class {\tt DataWidgetDef} contains the
            defining attributes from SQL table {\tt tb\_datawidget} (Fig.~\ref{fig:IVISIT_SQL_tables}) to allow creation of GUI widget (see {\tt ivisit.widgets}).
      \item {\tt CommentWidgetDef}: For each {\bf Comment Widget} from the GUI of an IVISIT application, a corresponding object from class {\tt CommentWidgetDef} contains the
        defining attributes from SQL table {\tt tb\_commentwidget} (Fig.~\ref{fig:IVISIT_SQL_tables}) to allow the creation of the GUI widget (see {\tt ivisit.widgets}).
      \end{itemize}
      In addition, there is a container class {\tt IVisitWidgetDefCollection} that holds {\it all} widget definitions for a simulation context defined in SQL table {\tt tb\_simulation}
      (Fig.~\ref{fig:IVISIT_SQL_tables}). The container class mirrors the records of the SQL tables, interchanges information with the GUI widgets,
      and saves modified information persistently in the SQL tables (e.g., after shifting a slider and pressing the save button).
\item {\bf ivisit.parser.py}: Provides class {\tt IVisitParser} for parsing ``in-code'' widget definitions from the Python code of an IVISIT application
      (see appendix~\ref{app:InCodeCommandsGUIWidgets} for an overview of in-code-commands). For each ``in-code'' command corresponding entries in the SQL tables
      are created (see Fig.~\ref{fig:IVISIT_SQL_tables} in appendix~\ref{app:UML_class_diagrams}).
      For example, parsing the command ``{\tt \#@IVISIT:SLIDER  \& name    \& [200,1] \& [0,9,3,1] \& var \& -1 \& int \& 0}''
      (from appendix~\ref{app:InCodeCommandsGUIWidgets}, page~\pageref{txt:item:SLIDER_directive}) will create 
      corresponding entries of type ``slider'' in the SQL tables {\tt tb\_parameter} and {\tt tb\_parameterwidget}. 
\item {\bf ivisit.defdb.py}: Defines the 6 SQL tables {\tt tb\_simulation}, {\tt tb\_parameter}, {\tt tb\_dataarray}, {\tt tb\_parameterwidget}, {\tt tb\_datawidget}, and {\tt tb\_comment widget}
      that persistently store the parameter and widget information of each simulation context of an IVISIT application (see Fig.~\ref{fig:IVISIT_SQL_tables} in appendix~\ref{app:UML_class_diagrams}).
\item {\bf ivisit.defdialogs.py}: Defines the dialogs for editing the SQL database tables via the menu item {\tt HUBS} and {\tt Databases} in an IVISIT application. With menu item {\tt Databases}
      the six SQL tables can be manipulated directly (cf., Fig.~\ref{fig:IVISIT_SQL_tables} in appendix~\ref{app:UML_class_diagrams}). By contrast, menu item {\tt HUBS} provides topical dialogs where
      multiple SQL tables are involved. For example, selecting menu item {\tt HUBS/Simulation} allows editing all simulation contexts and all corresponding parameter and data widgets.
\item {\bf ivisit.simulation.py}: Provides the base classes and templates to define an IVISIT simulation application. For example, each IVISIT simulation application
      should define a class like the template {\tt IVisit\_parameters}
      defining the simulation parameters to be controlled or manipulated by GUI elements and, similarly, a class like the template {\tt IVisit\_Data} to define simulation data to be displayed by GUI elements.
      And each IVISIT simulation application must be derived from the base class {\tt IVisit\_Simulation}. See section~\ref{sec:examples} for examples how to implement IVISIT simulation applications.
\end{itemize}

% *******************************************************************************************
% *******************************************************************************************
\section{Examples}  \label{sec:examples}
% *******************************************************************************************
% *******************************************************************************************

{\bf Installing IVISIT} requires a Python3 installation (at least version 3.6) \cite{PythonOrg:2023}, e.g., for
Microsoft Windows download from {\tt https://www.python.org/}, or for Linux/Ubuntu install by {\tt sudo apt install python3}.
It is also highly recommended to use virtual environments to separate different (incompatible) python versions \cite{Venv:2023}.
After that, install the python modules required for IVISIT using PIP \cite{PIP:2023}:
\begin{align*}
\mbox{{\tt pip install supylib}}\quad\quad\mbox{and}\quad\quad \mbox{{\tt pip install ivisit}}
\end{align*}
Besides, you may need to install a couple of further python packages, depending on your scope, for example:
{\tt pip install numpy},\ \ {\tt pip install matplotlib},\ \ {\tt pip install tkinter},\ \ {\tt pip install scipy},\ \ {\tt pip install opencv-python}, \ \ 
{\tt pip install scikit-image},\ \ {\tt pip install tensorflow}, \ \ {\tt pip install keras}, et cetera.

% *******************************************************************************************
\subsection{Example 1: Hello World}  \label{sec:example1_HelloWorld}
% *******************************************************************************************
After installing, you can write your first {\bf IVISIT simulation application}.
Listing~\ref{lst:HelloWorld} shows an (almost) minimal ``Hello World''
example that {\bf simulates a simple system} given by recursive definition
\begin{align}
   x(0):=100  \quad\quad\mbox{and}\quad\quad x(n):={\tt decay}\cdot x(n-1) \quad\mbox{for $n=1,2,\ldots$}  \label{eq:hello_world_system_recursion}
\end{align}
where {\tt decay}$\in\setR$ is a decay parameter (typically being $<1$).  
Type the text of {\bf listing~\ref{lst:HelloWorld}} into a {\bf text editor} (like Emacs or Notepad++) or copy \& paste it, and then save
it as {\tt demo01\_HelloWorld.py}. 

%\lstinputlisting[language={Python},caption={demo01\_HelloWorld.py},label={lst:HelloWorld}]{python/demo01_HelloWorld.py}
\begin{lstlisting}[language={Python},caption={demo01\_HelloWorld.py},label={lst:HelloWorld}]
#@IVISIT:SIMULATION & sim_HelloWorld1
#@IVISIT:SLIDER & Decay factor & [200,1] & [0.8,1.2,4,0.01] & decay & -1 & float & 0.9
#@IVISIT:SLIDER & Delay [msec] & [200,1] & [0  ,500,4,1]    & delay & -1 & int  &  100
#@IVISIT:TEXT_OUT & Results    & [20,5]  & just_left        & str_results 

import ivisit as iv
import time

# (I) define simulation parameters to be controlled by IVisit
class SimParameters(iv.IVisit_Parameters):
    decay=0.9            # decay factor for computing recursion x(n)=decay*x(n-1)
    delay=100            # delay in msec after each simulation step 
    
# (II) define simulation data to be displayed by IVisit
class SimData(iv.IVisit_Data):
    str_results='Hello World'  # string for text output
    
# (III) main class for simulation 
class Sim(iv.IVisit_Simulation):
    def __init__(self,name_arg="demo01_HelloWorld.py"): # constructor: called only once 
        iv.IVisit_Simulation.__init__(self,name_arg,SimParameters,SimData) # call base 
        self.init()      # call to init() to initialize state variable x

    def init(self):      # is called each time the "init" button is pressed  
        p,d     = SimParameters,SimData  # get short hands to parameters and data
        self.x = 100     # define initial value of variable x (for a simple recursion)
        d.str_results ="Hello World!\n"  # define initial textual output
        d.str_results+="Let us compute x(n) = "+str(p.decay)+"*x(n-1)\n"
        d.str_results+="using the initial value x(0)="+str(self.x)+"\n"
              
    def step(self):  # is called each simulation step (e.g., pressing "step" button) 
        p,d     = SimParameters,SimData  # get short hands to parameters and data
        self.x = self.x*p.decay  # update variable x by multiplying decay factor
        d.str_results="Hello World!\n"   # displaying results as textual output
        d.str_results+="Let us compute x(n) = "+str(p.decay)+"*x(n-1)\n"
        d.str_results+="Now we have x("+str(self.parent.steps)+")="+str(self.x)+"\n"    
        time.sleep(p.delay/1000) # sleep for a while to delay simulation progress
        
# (IV) main program
iv.IVisit_main(sim=Sim(),
               str_app_name='demo01_HelloWorld', # insert here your app name
               str_app_help='help text',         # use some more meaningful help text
               str_app_about='about text')       # use some text about your app
\end{lstlisting}

\noindent Here line 6 imports the IVISIT modules. Lines 1--4 define the simulation context {\tt sim\_HelloWorld1} and three GUI elements (two sliders and one text output field;
see appendix~\ref{app:InCodeCommandsGUIWidgets}).
The sliders control the simulation parameters {\tt decay} and {\tt delay} defined in lines 10--12 (within class {\tt SimParameters}),
whereas the text output field displays the string variable {\tt str\_results} defined in lines 15--16 (within class {\tt SimData}). 
Then the simulation is defined in lines 19--37 (within class {\tt Sim}):
The constructor (lines 20-22) defines the simulation name, initializes the base class {\tt IVisitSimulation} and calls {\tt init()}.
The {\tt init()} function (lines 24--29) initializes the system state variable {\tt x} (corresponding
to $x(0)$ in (\ref{eq:hello_world_system_recursion})) and the output string {\tt str\_results}. And the
{\tt step()} function (lines 31--37) computes on simulation time step by updating the system state {\tt x}
according to (\ref{eq:hello_world_system_recursion}), writing the text output {\tt str\_results}, and then sleeping for a time defined
by parameter {\tt delay}. Finally, the main program (lines 40-43) starts the IVISIT simulation application
by creating an object from class {\tt Sim()} and passing it to {\tt IVisit\_main(.)} along with strings
{\tt str\_app\_name}, {\tt str\_app\_help}, and {\tt str\_app\_about},
defining the application name (displayed as title in the app window), the ``help text'' displayed
when clicking on the help menu item, and the ``about text'' displayed when clicking in the ``about app''
menu item of the simulation application.

% Figure fig3_start_demo01_HelloWorld 
% created with PowerPoint on HS-Laptop
% --> see: Figur 3
%     in \Users\knoblauch\Desktop\Aktuelles_Draft\KEIM-Projekte\KICAD-CMCENgineers-hermle\Berichte\ivisit_techreport_materials\TechreportIVISIT_Figuren.pptx 
%     Kopie davon liegt auf: ~/hs-albsig/02-research/KICAD/ivisit_techreport/figures/ivisit_techreport_materials
% exportiert als png von PowerPoint; umgewandelt mit img2pdf in pdf
% mit ivisit snapshots in ~/hs-albsig/02-research/KICAD/ivisit_techreport/figures/snapshots:
%     fig_demo01_1.png - fig_demo01_6.png
\begin{figure}[th]
%\nopagenumber
%\renewcommand{\baselinestretch}{1.0}
\hfill
\begin{center}
\hspace*{-1cm}
\includegraphics[width=1.2\linewidth]{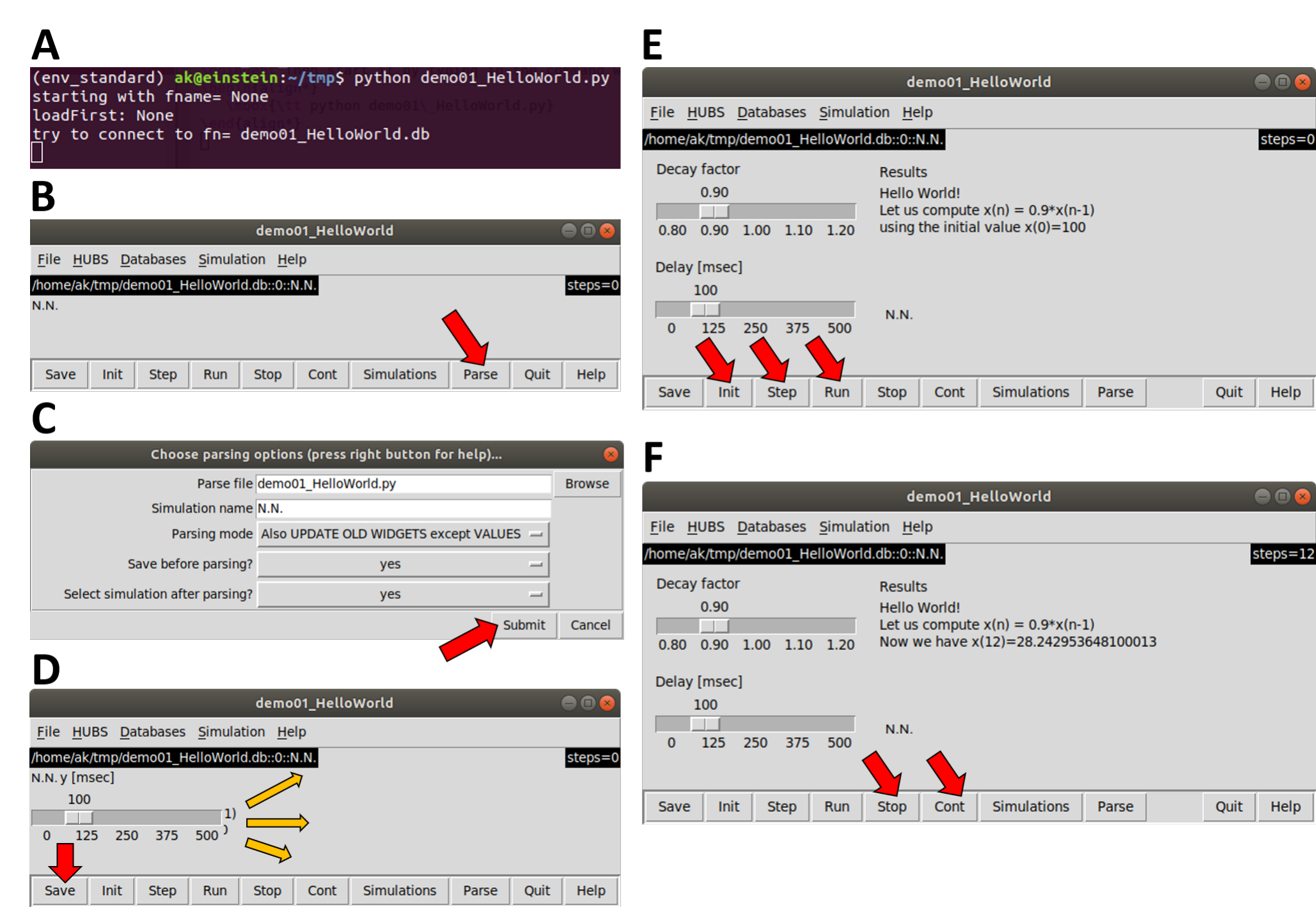}    
\end{center}
\caption{\label{fig:start_demo01_HelloWorld}
  Starting the IVISIT simulation application, parsing the GUI elements,
  and doing some simulation steps (see text for details).
} 
\end{figure}

Fig.~\ref{fig:start_demo01_HelloWorld}A shows how to 
{\bf start the IVISIT simulation application} of Listing~\ref{lst:HelloWorld}: After 
typing in the command window 
\begin{align*}
   \mbox{\tt python demo01\_HelloWorld.py}
\end{align*}
a window with the simulation application will open (Fig.~\ref{fig:start_demo01_HelloWorld}B).
On the bottom there are action buttons. Press the action button {\tt Parse}
to {\bf parse the definitions of the IVISIT GUI elements} (lines 1--4 in Listing~\ref{lst:HelloWorld}).
A dialog window opens where you can specify various parse options (Fig.~\ref{fig:start_demo01_HelloWorld}C).
Just confirm the default options by clicking the {\tt Submit} button. Now Listing~\ref{lst:HelloWorld} is parsed
and the corresponding GUI elements appear in the upper left area of the simulation application (Fig.~\ref{fig:start_demo01_HelloWorld}D). As the GUI elements are overlaying with each other, {\bf drag and drop the GUI elements}
to the location
where you want them to be (to drag and drop you have to click on the ``label'' of each GUI element, which is usually
in the top left area of the GUI element), and then press the {\tt Save} button to {\bf save the layout}.
Now you can {\bf initialize the simulation} by pressing the {\tt Init} button (which will call the
{\tt init()} function in Listing~\ref{lst:HelloWorld}), and {\bf do one or more simulation steps} by
clicking one or more times the
{\tt Step} button (Fig.~\ref{fig:start_demo01_HelloWorld}E).
This will call the {\tt step()} function in Listing~\ref{lst:HelloWorld} (lines 31--37), which multiplies
the system state {\tt x} by factor {\tt decay} according to (\ref{eq:hello_world_system_recursion}).
You can modify the parameter {\tt decay} by shifting the
slider named ``Decay Factor''. By clicking on the {\tt Run} button {\tt step()} will be called iteratively.
The number of simulation steps are indicated in the upper right corner (text field with black background).
If the simulation is too fast to follow the {\bf Results text field}, you may slow it down by increasing the parameter
{\tt Delay [msec]} by shifting the corresponding slider. To {\bf save the parameter settings} press the {\tt Save} button. To {\bf stop the simulation} press button {\tt Stop} (Fig.~\ref{fig:start_demo01_HelloWorld}D).
To {\bf continue the simulation} press button {\tt Cont}.
Finally, to {\bf quit the simulation} press the button {\tt Quit}.

Instead of pressing the action buttons you can equivalently select the corresponding {\bf menu bar items} (in the top area of the window): In the menu {\tt File} you can select {\tt Save}, {\tt Parse}, and {\tt Quit}. In the menu {\tt Simulation} you can select {\tt Init}, {\tt Step}, {\tt Run}, {\tt Stop}, and {\tt Cont}. In the menu {\tt help} you can get {\bf additional information and help on IVISIT and the simulation application}. There are also {\bf further menu functions}: For example, in the menu {\tt HUBS} and {\tt Databases} you can directly manipulate all attributes of the SQL tables by hand (see appendix~\ref{app:UML_class_diagrams}). With the menu items {\tt File/Open},
{\tt File/SaveAs}, {\tt File/New}, you can define new {\bf SQL database files} where the parameters
and GUI settings of the IVISIT simulation application are stored. Per default the SQL database file name is the same as the name of the simulation file, but ending with {\tt $\ast$.db}: For example, the default SQL database file name for
Listing~\ref{lst:HelloWorld} is {\tt demo01\_HelloWorld.db}. If you want to start a IVISIT simulation application with another SQL database file, you can pass this as an additional parameter in the command window. For example, typing  
\begin{align*}
   \mbox{\tt python demo01\_HelloWorld.py demo01\_parameters.db}
\end{align*}
to start the IVISIT simulation application of Listing~\ref{lst:HelloWorld}
will load (and save) the parameter and GUI settings from (to) the SQL database file
{\tt demo01\_parameters.db} instead of {\tt demo01\_HelloWorld.db}. Within each
IVISIT simulation application or SQL database file you can
select different {\bf simulation contexts} via menu item {\tt HUBS/Simulation}.
The default simulation context is {\tt N.N.} (you may change that to a more meaningful name), but you may add
alternative simulation contexts by clicking {\tt Veraendern} (=Modify) and then {\tt Neu} (=New) or
{\tt Kopieren} (=Copy). This will allocate new simulation contexts where you can store alternative sets of parameter values and/or GUI element settings, for example, corresponding to multiple/alternative optimal parameter sets.

% *******************************************************************************************
\subsection{Example 2: Simulating Leaky-Integrate\&Fire-Neurons using SimpleScope}  \label{sec:example2_LIFNeuron_simplescope}
% *******************************************************************************************

Next we simulate a simple leaky-integrate-and-fire (LIF-) neuron model \cite{Dayan/Abbott:2001,Gerstner/Kistler:2002,Knoblauch/Palm:2002_a,Knoblauch:2003_b}. A {\bf LIF-neuron} has a dendritic voltage $v(t)\in\setR$
and a spike variable $s(t)\in\{0,v_{\mathrm{spike}}\}$ following the differential equation
\begin{align}
   \tau v'(t)=-v(t)+I(t)  \quad\quad\mbox{and}\quad\quad s(t)=\begin{cases} v_{\mathrm{spike}}, & v(t)\ge \Theta \\ 0, & \mbox{else} \end{cases} \label{eq:LIF_neuron_DGL}
\end{align}
where $\tau$ is the dendritic membrane time constant, $I(t)$ is the input current, and $v(t_s+):=0$ is reset to voltage 0 each spike time $t_s$ with $s(t_s)=1$.
Here $I(t):=I_0+r(t)\sim\mathcal{N}(I_0,\sigma^2)$ is Gaussian with mean $I_0$ and
standard deviation $\sigma$. Listing~\ref{lst:LIFNeuron} shows a corresponding {\bf IVISIT simulation application}:

%\lstinputlisting[language={Python},caption={demo02\_LIFNeuron.py},label={lst:LIFNeuron}]{python/demo02_LIFNeuron.py}
\begin{lstlisting}[language={Python},caption={demo02\_LIFNeuron.py},label={lst:LIFNeuron}]
#@IVISIT:SIMULATION & sim_LIFNeuron1 
#@IVISIT:SLIDER & input current I0 & [200,1] & [-10,30,4,0.1]& I0     & -1 & float & 2
#@IVISIT:SLIDER & noise s.d. sigma & [200,1] & [0,100,4,0.5] & sigma  & -1 & float & 1
#@IVISIT:SLIDER & threshold theta  & [200,1] & [1,100,4,0.5] & theta & -1 & float & 10 
#@IVISIT:SLIDER & time step size dt &[200,1] & [0.01,10,4,0.01] & dt & -1 & float & 1
#@IVISIT:IMAGE    & dendritic voltage v & 1.0 & [0,255] & im_voltage & int
#@IVISIT:TEXT_OUT & Results & [20,5] & just_left & str_info 

import numpy as np
import ivisit as iv
from ivisit.special import SimpleScope

# (I) define parameters to be controlled by IVisit
class SimParameters(iv.IVisit_Parameters):
    tau   =20       # membrane time constant of neuron [msec]
    theta =10       # firing threshold [mV]
    I0     =2       # (mean) input current [mA]
    sigma =0.1      # standard deviation of Gaussian noise on I [mA]
    dt    =0.1      # simulation time step size [msec]
    v_spike=30      # voltage of a spike [mV]
    t_min,t_max,nt=0,1000,400  # range of time axis (min, max, number of pixels)
    v_min,v_max,nv=-10,40,200  # range of voltage axis (min, max, number of pixels)
    
# (II) define simulation data to be displayed by IVisit
class SimData(iv.IVisit_Data):
    im_voltage  = np.array([[0]]) # image for displaying membrane voltage of neuron
    str_info    = ''              # string for text output
    
# (III) main class for simulation 
class Sim(iv.IVisit_Simulation):
    def __init__(self,name_arg="demo02_LIFNeuron.py"):
        iv.IVisit_Simulation.__init__(self,name_arg,SimParameters,SimData)
        self.scope_v=SimpleScope() # define simple scope to display neuron recordings
        self.init()

    def init(self):
        p,d = SimParameters,SimData # short hands to simulation parameters and data
        d.im_voltage=self.scope_v.configure(p.nt,p.nv,p.t_min,p.t_max,p.v_min,p.v_max)    # configure scope to display neuron recordings in im_voltage
        self.v,self.s = 0,0         # initialize dendritic voltage and spike variables
        
    def step(self):
        # (i) update neuron states (solve LIF differential equation tau*v'=-v+I+noise)
        p,d     = SimParameters,SimData  # short hands to parameters and data
        v_old   = self.v                 # old voltage from previous simulation step
        t       = self.parent.steps*p.dt # current simulation time
        I       = p.I0+p.sigma*np.random.randn(1)[0] # input current with Gauss noise
        self.v  = self.v + (I-self.v)/p.tau*p.dt  # Euler integration of diff.eq.
        self.scope_v.set_data(t,v_old,self.v,128) # draw voltage in image (gray)
        self.s  = 0                               # default: there is no spike
        if self.v>=p.theta:                       # if v exceeds firing threshold ?
            self.s,self.v=p.v_spike,0                     # then spike and reset v
            self.scope_v.set_data(t,self.v,p.v_spike,255) # and draw spike (white)
        
        # (ii) display simulation results also as text 
        d.str_info ="time    t="+str(round(t,3))+"\n"          
        d.str_info ="input   I="+str(round(I,3))+"\n"          
        d.str_info+="voltage v="+str(round(self.v,3))+" v_old="+str(round(v_old,3))
        d.str_info+="\nspike   s="+str(self.s)+"\n"
        
# (IV) main program
iv.IVisit_main(sim=Sim(),str_app_name='demo02_LIFNeuron')
\end{lstlisting}

% Figure fig_demo02_LIFNeuron 
% created with the Ubuntu Bildschirmfoto-Tool and saved as png in ~/hs-albsig/02-research/KICAD/ivisit_techreport/figures/snapshots
% converted to pdf using img2pdf in the Windows operation system
\begin{figure}[hb]
%\nopagenumber
%\renewcommand{\baselinestretch}{1.0}
\hfill
\begin{center}
\includegraphics[width=\linewidth]{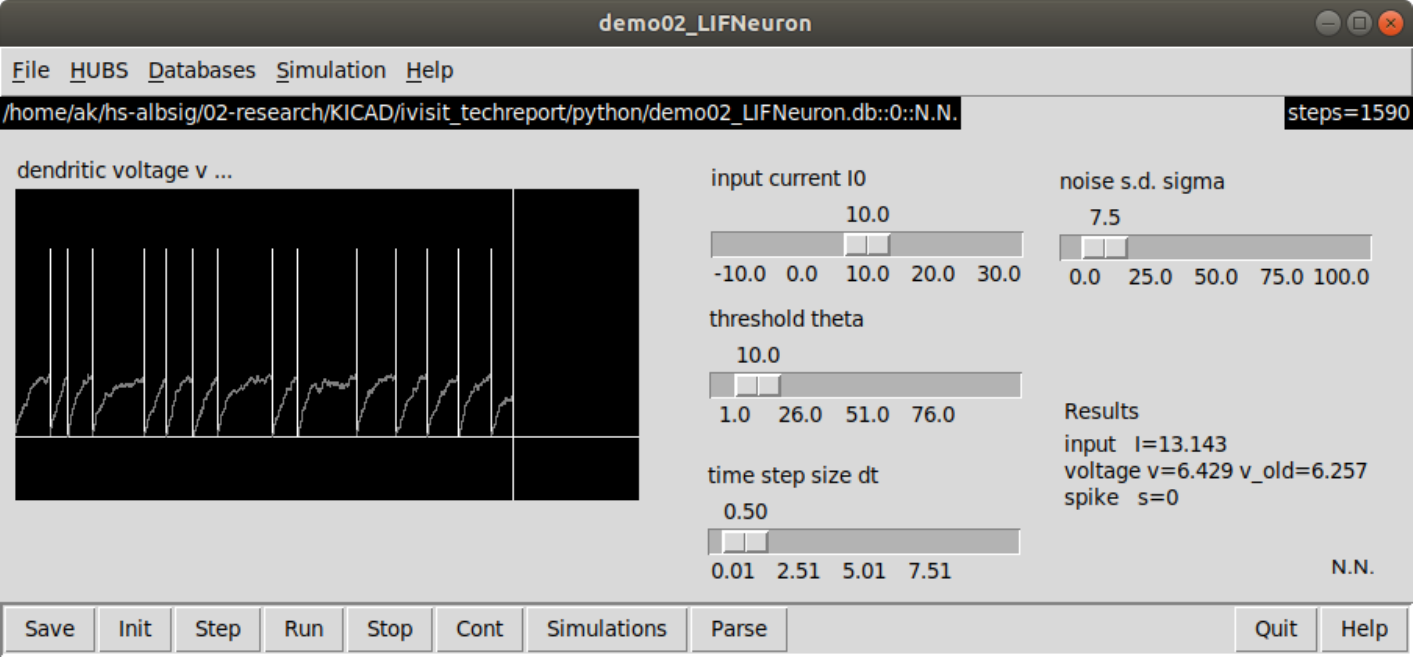}    
\end{center}
\caption{\label{fig:run_demo02_LIFNeuron}
  Running the IVISIT application for simulating a LIF-Neuron (Listing~\ref{lst:LIFNeuron}).
  The dendritic voltage trace $v$ of the LIF-Neuron from (\ref{eq:LIF_neuron_DGL},\ref{eq:LIF_neuron_DGL_discretized})
  is drawn in a {\tt SimpleScope} imported from {\tt ivisit.special} and displayed within a corresponding
  IMAGE widget.
} 
\end{figure}

\noindent The general structure of listing~\ref{lst:LIFNeuron} is the same as in Example 1 (see Listing~\ref{lst:HelloWorld}):
Lines 1--7 define the {\bf IVISIT GUI elements}, including 4 sliders to control the neuron parameters $I_0, \sigma, \theta$, and the simulation step size
$\Delta t$ via the corresponding Python variables {\tt I0}, {\tt sigma}, {\tt theta}, and {\tt dt}. Line 8 defines an IMAGE to display the voltage trace of the neuron,
which is drawn to a 2D Numpy array {\tt im\_voltage}. Lines 14--22 define the {\bf simulation parameters}. Note that there may be additional parameters like {\tt tau}, {\tt v\_spike}, and the range parameters {\tt t\_min}, {\tt t\_max}, \ldots, {\tt nv}, that need not be controlled by any GUI element. Any parameter that is not controlled by a GUI element (like a slider) will just keep its default value as defined in class {\tt SimParameters}. Lines 25--27 define the {\bf simulation data}, which includes the 2D image array {\tt im\_voltage}
and a string variable {\tt str\_info} for text output. Note that {\tt im\_voltage} can be initialized here with minimal size $1\times 1$, as it will be allocated to the correct size in {\tt init()} (line 38). 
Lines 30--58 defines the {\bf simulation class} including {\tt init()} and {\tt step()}. Line 38 of {\tt init()} defines a {\tt SimpleScope} object
(imported from {\tt ivisit.special.py}; see section~\ref{sec:architecture}, page~\pageref{txt:item:ivisist_special})
for displaying the neuron's voltage trace, which
includes a 2D image array of the desired size that is returned to the simulation data array {\tt im\_voltage}. In {\tt step()}, line 47 applies Euler integration of the differential equation (\ref{eq:LIF_neuron_DGL}).
That is, the derivative $v'(t)\approx \frac{v(t)-v(t-\Delta t)}{\Delta t}$ is approximated by its difference quotient such that (\ref{eq:LIF_neuron_DGL}) becomes
$\tau\frac{v(t)-v(t-\Delta t)}{\Delta t}\approx -v(t-dt)+I(t)$, and
solving for $v(t)$ yields equivalently
\begin{align}
   v(t)\approx v(t-\Delta t) + \frac{I(t)-v(t-\Delta t)}{\tau}\cdot\Delta t \label{eq:LIF_neuron_DGL_discretized}
\end{align}
as implemented in line 47. Then line 48 draws the new data point $(t,v(t))$ on the {\tt SimpleScope} (and thus also to the simulation data array {\tt im\_voltage}).
Here {\tt v\_old} corresponds to the ``old'' voltage $v(t-\Delta t)$ of the previous simulation step, and is also passed to the {\tt SimpleScope} to connect the old voltage $v(t-\Delta t)$ with the new voltage $v(t)$
by a line having gray value 128. Lines 49--52 handle the spike variable. If $v(t)\ge\theta$ (line 50) the spike variable {\tt s} is set to the spike voltage $v_{\mathrm{spike}}$ and also displayed on the {\tt SimpleScope}
(line 52). Lines 55--58 write the text output to {\tt str\_info}, and line 61 starts the main program, all similar to Listing~\ref{lst:HelloWorld}. After {\bf starting}, parsing, and drag\&drop-ing the GUI elements
similar as in example~1 (section~\ref{sec:example1_HelloWorld}, Fig.~\ref{fig:start_demo01_HelloWorld}), the simulation window should look similar to Fig.~\ref{fig:run_demo02_LIFNeuron}. 

You can {\bf save the IMAGE widgets} at any time to an image file in the usual image formats (jpg, png, tif, eps etc.):
If you click on the the three dots {\tt ...} on the right side of the label (here {\tt dendritic voltage v}) then a menu window opens,
and you can select menu item {\tt Save image} opening a file dialog. There are also {\bf further menu options},
for example, selecting specific image channels, scaling the image, accessing properties of the image widget, or hiding the image.
Click again on the three dots (at the same position) to close the menu (if you have moved the mouse, just click two times on the three dots).

% *******************************************************************************************
\subsection{Example 3: Simulating Leaky-Integrate\&Fire-Neurons using Matplotlib}  \label{sec:example3_LIFNeuron_matplotlib}
% *******************************************************************************************
To get {\bf high quality plots} similar to Matlab \cite{MATLAB:2022} you may use the {\bf Matplotlib library} \cite{Hunter:Matplotlib:2007} by importing
IVISIT package {\tt ivisit.matplotlib} (see section~\ref{sec:architecture}, page~\ref{txt:item:ivisist_matplotlib}). 
The following Listing~\ref{lst:LIFNeuron_matplotlib} shows a simulation of a LIF neuron similar to the previous example (section~\ref{sec:example2_LIFNeuron_simplescope}),
but using Matplotlib instead of {\tt ivisit.special.SimpleScope} to display the neuron's voltage trace and spikes: 

%\lstinputlisting[language={Python},caption={demo03\_LIFNeuron\_matplotlib.py},label={lst:LIFNeuron_matplotlib}]{python/demo03_LIFNeuron_matplotlib.py}
\begin{lstlisting}[language={Python},caption={demo03\_LIFNeuron\_matplotlib.py},label={lst:LIFNeuron_matplotlib}]
#@IVISIT:SIMULATION & sim_LIFNeuron1_matplotlib 

#@IVISIT:DICTSLIDER & Neuron parameters & [200,30,-1,2,10] & par_neuron & 2
#@IVISIT:DICTSLIDERITEM & time constant tau   & [1,100,3,0.5] & tau   & float & 20
#@IVISIT:DICTSLIDERITEM & threshold theta     & [1,100,3,0.5] & theta & float & 10 
#@IVISIT:DICTSLIDERITEM & input current I0    & [-10,30,4,0.1]& I0    & float &  2
#@IVISIT:DICTSLIDERITEM & noise s.d. sigma    & [0,100,4,0.5] & sigma & float &  1
#@IVISIT:DICTSLIDERITEM & spike voltage v_spk & [0,200,4,1] & v_spike & float & 30
#@IVISIT:DICTSLIDERITEM & time step size dt   & [0.01,10,4,0.01] & dt & float &  1

#@IVISIT:DICTSLIDER & Display parameters & [200,30,-1,2,10] & par_display & 0
#@IVISIT:DICTSLIDERITEM & disp_skip & [1,100,4,1]  & disp_skip & int   &    1
#@IVISIT:DICTSLIDERITEM & t_tail    & [1,2000,4,1] & t_tail    & int   & 1000
#@IVISIT:DICTSLIDERITEM & t_head    & [1,100,4,1]  & t_head    & int   &   10
#@IVISIT:DICTSLIDERITEM & v_min     & [-50,50,4,1] & v_min     & float &  -10
#@IVISIT:DICTSLIDERITEM & v_max     & [10,200,4,1] & v_max     & float &   40

#@IVISIT:IMAGE    & dendritic voltage v & 1.0 & [0,255] & im_voltage & int
#@IVISIT:TEXT_OUT & Results & [20,5] & just_left & str_info 

import numpy as np
import ivisit as iv
from ivisit.matplotlib import getMatplotlibFigure, getMatplotlibImage

# (I) define parameters to be controlled by IVisit
class SimParameters(iv.IVisit_Parameters):
    par_neuron = {
        'tau'    :20   , # membrane time constant of neuron [msec]
        'theta'  :10   , # firing threshold [mV]
        'I0'     : 2   , # (mean) input current [mA]
        'sigma'  : 0.1 , # standard deviation of Gaussian noise on I [mA]
        'dt'     : 0.01, # simulation step size [msec]
        'v_spike':30     # voltage of a spike [mV]
    }
    par_display = {
        'disp_skip':1,   # do matplotlib plot only every disp_skip steps
        't_tail':1000,   # plot voltage/spikes for last t_tail milliseconds [msec]
        't_head':  10,   # plot time t_head ahead of presence (padding) [msec]
        'v_min' : -10,   # minimum of voltage range [mV]
        'v_max' :  40    # maximum of voltage range [mV}
    }
        
# (II) define simulation data to be displayed by IVisit
class SimData(iv.IVisit_Data):
    im_voltage = np.array([[0]]) # image for displaying membrane voltage of neuron
    str_info   = ''              # string for text output
    
# (III) main class for simulation 
class Sim(iv.IVisit_Simulation):
    def __init__(self,name_arg="demo03_LIFNeuron_matplotlib.py"):
        iv.IVisit_Simulation.__init__(self,name_arg,SimParameters,SimData)
        self.init()

    def init(self):
        self.t, self.v, self.s = 0,0,0  # initialize time, voltage, spike variables
        self.t_rec, self.v_rec, self.s_rec = [0],[0],[0] # and recording lists 
        
    def step(self):
        # (i) update neuron states (solve LIF differential equation tau*v'=-v+I+noise)
        pn,pd = SimParameters.par_neuron, SimParameters.par_display # shortcut dicts
        d = SimData                                                 # and sim. data 
        self.t  = self.t+pn['dt']                      # increase time by dt
        I_noisy = pn['I0']+pn['sigma']*np.random.randn(1)[0] # noisy input current 
        self.v  = self.v + (I_noisy-self.v)/pn['tau']*pn['dt'] # Euler integration 
        if self.v>=pn['theta']: self.s,self.v=pn['v_spike'],0  # above thresh.? Spike! 
        else              : self.s       =0            # else no spike 
        self.t_rec.append(self.t)                      # append time to recordings
        self.v_rec.append(self.v)                      # append voltage to recordings 
        self.s_rec.append(self.s)                      # append spike to recordings 

        # (ii) textual outputs
        d.str_info ="time    t="+str(round(self.t,3))+"\n"
        d.str_info+="voltage v="+str(round(self.v,3))+"\n"
        d.str_info+="spike   s="+str(round(self.s,3))+"\n"

        # (iii) plot neuron recordings using matplotlib
        if self.parent.steps%pd['disp_skip']==0:          # skip matplotlib plot?
            self.fig = getMatplotlibFigure(figsize=(6,5)) # create figure
            ax1 = self.fig.add_subplot(111)               # create axis for plot 
            ax1.plot(self.t_rec,self.s_rec,'r')           # plot spikes 
            ax1.plot(self.t_rec,self.v_rec,'b')           # plot dendritic voltages 
            ax1.set_xlim((self.t-pd['t_tail'],self.t+pd['t_head'])) # limits time axis
            ax1.set_ylim((pd['v_min'],pd['v_max']))       # limits for y/voltage axis
            ax1.set_xlabel('time [msec]')                 # label for time-axis
            ax1.set_ylabel('dendritic voltage [mV]')      # label for voltage-axis            
            d.im_voltage = getMatplotlibImage(self.fig)   # send plot to ivisit image
        
# main program
iv.IVisit_main(sim=Sim(),str_app_name='demo03_LIFNeuron_matplotlib')
\end{lstlisting}

\begin{figure}[th]
%\nopagenumber
%\renewcommand{\baselinestretch}{1.0}
\hfill
\begin{center}
\includegraphics[width=\linewidth]{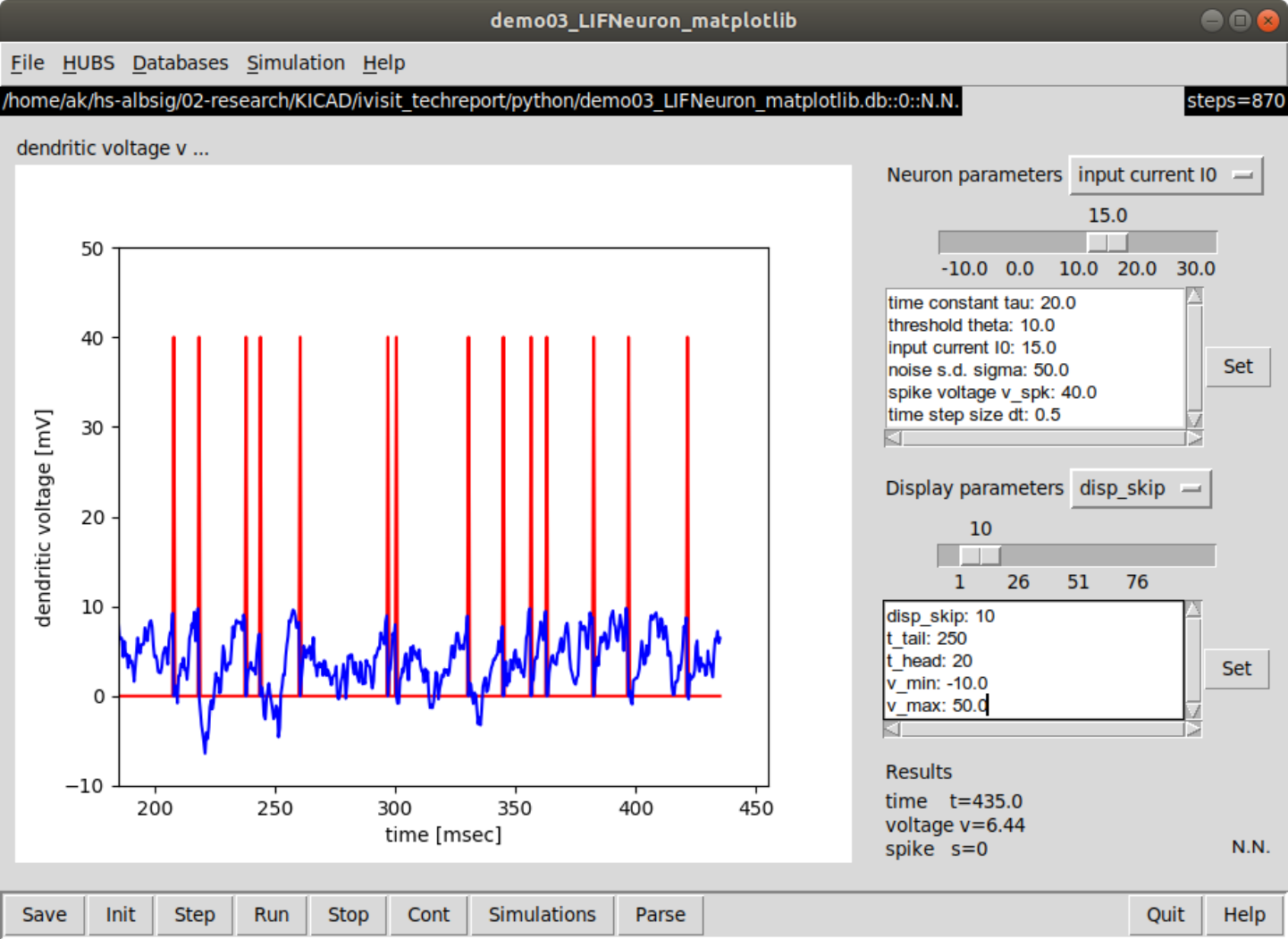}    
\end{center}
\caption{\label{fig:run_demo03_LIFNeuron_matplotlib}
  Running the IVISIT application for simulating a LIF-Neuron and plotting voltage traces and spikes
  using Matplotlib (Listing~\ref{lst:LIFNeuron_matplotlib}). 
  Simulation parameters are collected in DICT SLIDERS. Compare to Fig.~\ref{fig:run_demo02_LIFNeuron}.
} 
\end{figure}

To {\bf display Matplotlib plots within an IVISIT simulation application} you first have to create a Matplotlib figure object
by calling {\tt ivisit.matplotlib.getMatplotlibFigure()} (line 78).
With this you can do anything you can do with Matplotlib. For example, line 79 creates an axis object holding a single subplot.
Then lines 80--85 plot the voltage trace, the spikes, set limits for x-axis (corresponding to time) and y-axis (corresponding to voltage)
and labels for the x/y-axis. Finally you have to extract the image array from the Matplotlib figure by calling 
{\tt ivisit.matplotlib.getMatplotlibImage()} and assign it to the IVISIT image data array (line 86). Plotting with {\bf Matplot may be slower} than
directly storing simulation data in Numpy image arrays (as done in the previous example (section~\ref{sec:example2_LIFNeuron_simplescope}).
To {\bf accelerate the simulation} you may do the Matplotlib plotting not in each simulation step, but instead only in every 2nd, 3rd, or 10th step, for example.
This is realized in line 77 by an IF-clause checking (employing the modulo-operator {\tt \%}
whether the simulation step number is a multiple of the extra parameter {\tt disp\_skip}. By this the plots are updated only once every {\tt disp\_skip}
simulation steps. Another novel feature are the {\bf DICT SLIDERS} defined in lines 3--16 (cf., appendix~\ref{app:InCodeCommandsGUIWidgets}).
They allow to manipulate an arbitrary number of numeric simulation parameters using a slider (or a text field). For this the simulation
parameters have to be defined as Python dict items (lines 27--41). DICTSLIDERS are useful, in particular, if you want to manipulate {\bf many parameters}
and avoid the cluttering of your IVISIT simulation application with lots of individual SLIDER widgets. Fig.~\ref{fig:run_demo03_LIFNeuron_matplotlib} shows a snapshot from running
Listing~\ref{fig:run_demo03_LIFNeuron_matplotlib}. 

% Figure fig_demo03_LIFNeuron_matplotlib 
% created with the Ubuntu Bildschirmfoto-Tool and saved as png in ~/hs-albsig/02-research/KICAD/ivisit_techreport/figures/snapshots
% converted to pdf using img2pdf in the Windows operation system

% *******************************************************************************************
\subsection{Example 4: Data generation with interactive GUI elements and Matplotlib}  \label{sec:example4_InteractiveMatplotlibGUI}
% *******************************************************************************************
It is often desirable to interact with figures created by Matplotlib, which use a different coordinate system than the
IVISIT IMAGE widgets based on Tkinter. 
The following Listing~\ref{lst:InteractiveDataGeneration} shows a corresponding example of an IVISIT simulation application having {\bf interactive GUI elements} with
{\bf Matplotlib Figures} using {\bf Tkinter events}, where data points can be added, deleted or moved by mouse clicks and drag\&drop:

%\lstinputlisting[language={Python},caption={demo04\_InteractiveDataGeneration.py},label={lst:InteractiveDataGeneration}]{python/demo04_InteractiveDataGeneration.py}
\begin{lstlisting}[language={Python},caption={demo04\_InteractiveDataGeneration.py},label={lst:InteractiveDataGeneration}]
import threading # required for locks (to ensure mutual exclusive data access)
import numpy as np
import ivisit as iv
from ivisit.matplotlib import * 

#@IVISIT:SIMULATION & demo04_InteractiveDataCreation1
#@IVISIT:LISTSEL & Action & [10,3] & [New,Delete,Move] & str_action &-1& string & New
#@IVISIT:IMAGE    & Data    & 1.0    & [0,255]   & im_data     & int
#@IVISIT:TEXT_OUT & Output Information & [20,5] & just_left & str_info

def getNearestNeighbor(X,x): # return index of nearest neighbor of x in X[1],X[2],...
    dist = [np.linalg.norm(xi-x) for xi in X]  # list of distances between x and X[i] 
    return np.argmin(dist)   # return index idxNN of nearest neighbor X[idxNN] 
    
# (I) define parameters to be controlled by IVisit
class SimParameters(iv.IVisit_Parameters):
    str_action               = 'New'      # data action mode (New, Delete, Move)
    x1min,x1max, x2min,x2max = -8,8, -8,8 # axis limits
    
# (II) define simulation data to be displayed by IVisit
class SimData(iv.IVisit_Data):
    str_info = ''                     # string for text output
    im_data  = np.array([[0,0,0]])    # image of data vectors

# (III) main class for simulation 
class Sim(iv.IVisit_Simulation):
    def __init__(self,name_arg="demo04_InteractiveDataGeneration.py",flagLock=True):
        iv.IVisit_Simulation.__init__(self,name_arg,SimParameters,SimData)
        self.data_canvas=None         # image canvas
        self.ax1=None                 # reference to matplotlib axis
        self.init()                   # call init()
        
    def init(self):
        self.X=np.array([[2,1],[2,2],[2,-1]],'float') # base data set 
        self.last_pos = None          # data point clicked last
        self.idx_NN = None            # index of nearest neighbor data point
        self.state='IDLE'             # mouse event state
        
    def bind(self,parent,display): # parent=ivisit object (having display to bind to)
        # bind TKinter mouse events to canvas of IMAGE widget "Data" defined above
        display.bind2Widget("Data","<Button-1>",self.onPressedB1_data,"imgcanvas")
        display.bind2Widget("Data","<B1-Motion>",self.onMovedB1_data,"imgcanvas")
        display.bind2Widget("Data","<ButtonRelease-1>",self.onRelB1_data,"imgcanvas")

    def onPressedB1_data(self,event): # when clicking mouse button B1 on data widget
        if self.state=='IDLE':    # click in data IMAGE only relevant in state IDLE
            self.lock.acquire()   # make mouse processing thread-safe
            p,d=SimParameters,SimData # short hand to simulation parameters and data
            x,y=getDataPos_from_PixelPos(event.x,event.y,d.im_data.shape[0],self.ax1)
            self.last_pos=[x,y]   # store clicked position (in Matplotlib coordinates)
            if p.str_action=='New':   # if action mode is 'New' ?
                self.X=np.concatenate((self.X,[self.last_pos])) # add new data point
            elif len(self.X)>0:   # 'Delete' or 'Move' only relevant if there is data
                self.idx_NN=getNearestNeighbor(self.X,np.array([x,y])) # get index NN
                if p.str_action=='Delete':  # if action mode is 'Delete' then...
                    self.X=np.delete(self.X,self.idx_NN,axis=0) # ...delete NN data 
                else:                       # if action mode is 'Move' then... 
                    self.state='MOVING'     # switch state from IDLE to MOVING
                    self.X[self.idx_NN,:]=self.last_pos # update position of NN data
            self.lock.release()   # release lock
        
    def onMovedB1_data(self,event): # called when moved mouse button B1 on data widget
        p,d=SimParameters,SimData   # short hand to simulation parameters and data
        if p.str_action=='Move' and self.state=='MOVING': # relevant in mode MOVING
            self.lock.acquire()   # make processing of mouse events thread-safe
            x,y=getDataPos_from_PixelPos(event.x,event.y,d.im_data.shape[0],self.ax1)
            self.last_pos=[x,y]   # store clicked position (in Matplotlib coordinates)
            self.X[self.idx_NN,:]=self.last_pos # move position of data point
            self.lock.release()   # release lock
        
    def onRelB1_data(self,event): # when releasing mouse button B1 on data widget
        self.state='IDLE'         # just return to state idle
        
    def step(self):
        # (i) write some infos about the data in string
        p,d=SimParameters,SimData # short hand to simulation parameters and data
        d.str_info="-------------------------------------\n"
        d.str_info+="number of data points N="+str(len(self.X))+"\n"
        d.str_info+="str_action="+str(p.str_action)+"\n"
        d.str_info+="state="+str(self.state)+"\n"
        d.str_info+="last_pos="+str(self.last_pos)+"\n"
        d.str_info+="idx_NN="+str(self.idx_NN)+"\n"
        
        # (ii) plot data
        fig = getMatplotlibFigure(figsize=(5,4))            # create figure
        ax1 = fig.add_subplot(111)                          # create axis for plot 
        ax1.scatter(self.X[:,0],self.X[:,1],c='r',marker='x',s=40) # plot data
        ax1.set_xlabel('x1'); ax1.set_ylabel('x2')          # labels on x/y-axis
        ax1.grid(which='both')                              # draw a grid
        ax1.set_xlim((p.x1min,p.x1max))                     # limits for axis
        ax1.set_ylim((p.x2min,p.x2max))                     # limits for axis
        d.im_data = getMatplotlibImage(fig) # send matplotlib image to ivisit image
        self.ax1=ax1 # required for ivisit.matplotlib.getDataPos_from_PixelPos(.)
        return
               
# (IV) main program
iv.IVisit_main(sim=Sim(),str_app_name='demo04_InteractiveDataGeneration')
\end{lstlisting}

% Figure fig_demo04_InteractiveDataGeneration 
% created with the Ubuntu Bildschirmfoto-Tool and saved as png in ~/hs-albsig/02-research/KICAD/ivisit_techreport/figures/snapshots
% converted to pdf using img2pdf in the Windows operation system
\begin{figure}[th]
%\nopagenumber
%\renewcommand{\baselinestretch}{1.0}
\hfill
\begin{center}
\includegraphics[width=\linewidth]{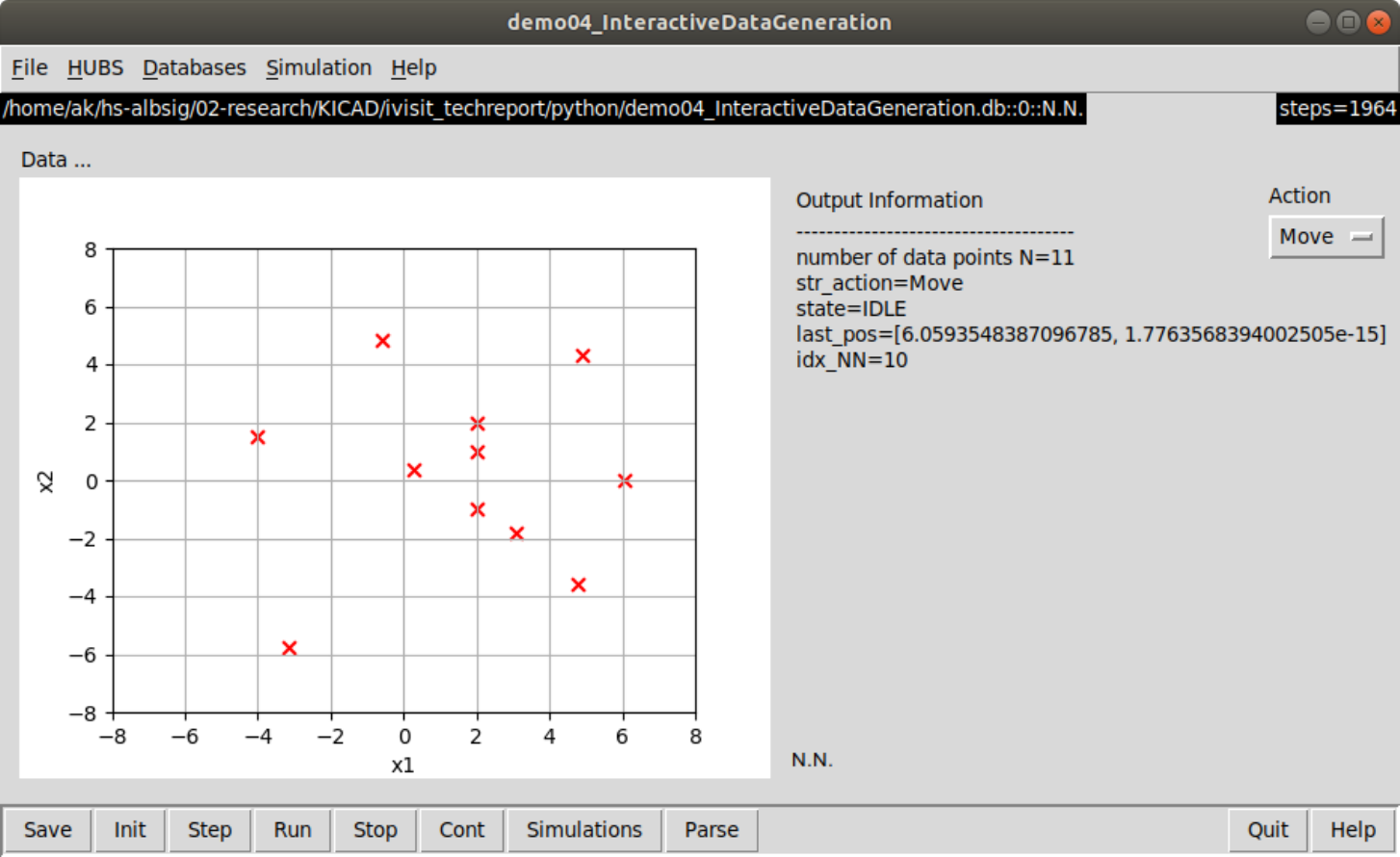}    
\end{center}
\caption{\label{fig:run_demo04_InteractiveDataGeneration}
  Running the IVISIT application for interactively generating (or deleting or moving)
  data by clicking on a Matplotlib figure (Listing~\ref{lst:InteractiveDataGeneration}).
  The action (New, Delete, or Move) can be chosen from the LISTSEL widget ``Action'' (upper right). 
} 
\end{figure}

The {\bf novel elements} of Listing~\ref{lst:InteractiveDataGeneration} are the {\tt bind()} function and the mouse event functions {\tt onPres\-sedB1\_data(.)}, {\tt onMovedB1\_data(.)},
and {\tt onRelB1\_data(.)} of class {\tt Sim} (lines 39--72). The function {\tt bind(parent,display)} can be used to {\bf bind event handling functions to TKinter events}
like {\tt <Button-1>}, {\tt <B1-Motion>}, and {\tt <ButtonRelease-1>} (lines 39--43). Similar to {\tt init()}, the function {\tt bind(parent,display)}
is called by the parent IVISIT simulation object's {\tt onSim\_init()} function,
but only after the display object (of type {\tt IVisitRawDisplayFram} defined in {\tt ivisit.widgets.py}) has been created, containing the TKinter widgets (by contrast, {\tt init()} is called already before
by {\tt onSim\_main\_init()}). So here {\tt onPressedB1\_data(event)} is called each time when the {\bf left mouse button is clicked on the IMAGE widget} (lines 45--60). Similarly,
{\tt onMovedB1\_data(event)} is called when the left button has been pressed and the mouse is moved over the IMAGE widget (lines 62--69), and
{\tt onRelB1\_data(event)} is called when the left mouse button is released again (lines 71--72). Together, these functions implement a {\bf state automaton} with two states {\tt IDLE} and {\tt MOVING}.
First, the state is initialized with {\tt self.state='IDLE'} in line~37. Then the action of a click on the IMAGE widget depends on the {\bf action mode} {\tt SimParameters.str\_\-action} (see lines 17 and 7):
For {\tt str\_action='New'} a {\bf new data point} will be added to the data array {\tt self.X} (initialized in line~37).
For this, line~49 calls {\tt ivisit.matplotlib.getDataPos\_\-from\_PixelPos(event.x,event.y,d.im\_data.shape[0],self.ax1)} in order to transform the
clicked canvas pixel position (from the Tkinter {\tt event})
to Matplotlib position {\tt [x,y]} (of the axis object {\tt self.ax1})
stored in {\tt self.last\_pos} (line 50), and to concatenate this to the data array {\tt self.X} (line 52). Similarly, {\tt str\_action='Delete'} will {\bf delete the data point}
that is closest to the clicked position (line 56), where function {\tt getNearestNeighbor(X,x)} (lines 11--13) is used to compute the ``nearest-neighbor'' (=closest point)
of the stored data {\tt X=self.X} to the clicked position {\tt x=self.last\_pos}. A more complex action is to {\bf move a data point} for {\tt str\_action='Move'}: Here the position of the ``nearest-neighbor'' data point
is updated with the clicked position (line 59) and the state of the automaton is switched to {\tt self.state='MOVING'} (line 58). Each time the mouse is moved (while still pressing the left button),
function {\tt onMovedB1\_data(event)} is called and correspondingly updates the position of the data point (line 68), until the left mouse button is released, which will
reset the state to {\tt self.state='IDLE'} in line~72 of function {\tt onRelB1\_data(event)}. Note that {\bf locks} are used to ensure mutual exclusive access to the data array {\tt self.X} and {\tt self.last\_pos}
(see lines 1, 27, 47 and 60, and 65 and 69). This is necessary as the mouse events are handled by another thread than the simulation main loop, and without the locks inconsistent behavior may occur
if one thread overwrites data read by the other thread. By setting {\tt flagLock=True} in line~27, the
simulation methods {\tt init(.)}, {\tt bind(.)}, and {\tt step(.)} are protected automatically against concurrent data access.
By contrast, each mouse event handling method like {\tt onPressedB1\_data(.)} and {\tt onMovedB1\_data(.)} has to be protected explicitly by hand, by bracketing the relevant code by {\tt self.lock.acquire()} and {\tt self.lock.release()} (see lines 47 and 60, and 65 and 69). Fig.~\ref{fig:run_demo04_InteractiveDataGeneration} shows a snapshot from running
Listing~\ref{lst:InteractiveDataGeneration}.

% *******************************************************************************************
\subsection{Example 5: Interactive classification using IVISIT event automatons}  \label{sec:example5_InteractiveGUI_IVISIT_eventautomatons}
% *******************************************************************************************

% Figure fig4_ClickDragEventAutomaton
% created with PowerPoint on HS-Laptop
% --> see: Figur 4
%     in \Users\knoblauch\Desktop\Aktuelles_Draft\KEIM-Projekte\KICAD-CMCENgineers-hermle\Berichte\ivisit_techreport_materials\TechreportIVISIT_Figuren.pptx 
%     Kopie davon liegt auf: ~/hs-albsig/02-research/KICAD/ivisit_techreport/figures/ivisit_techreport_materials
% exportiert als png von PowerPoint; umgewandelt mit img2pdf in pdf
\begin{figure}[b]
%\nopagenumber
%\renewcommand{\baselinestretch}{1.0}
\hfill
\begin{center}
\includegraphics[width=0.75\linewidth]{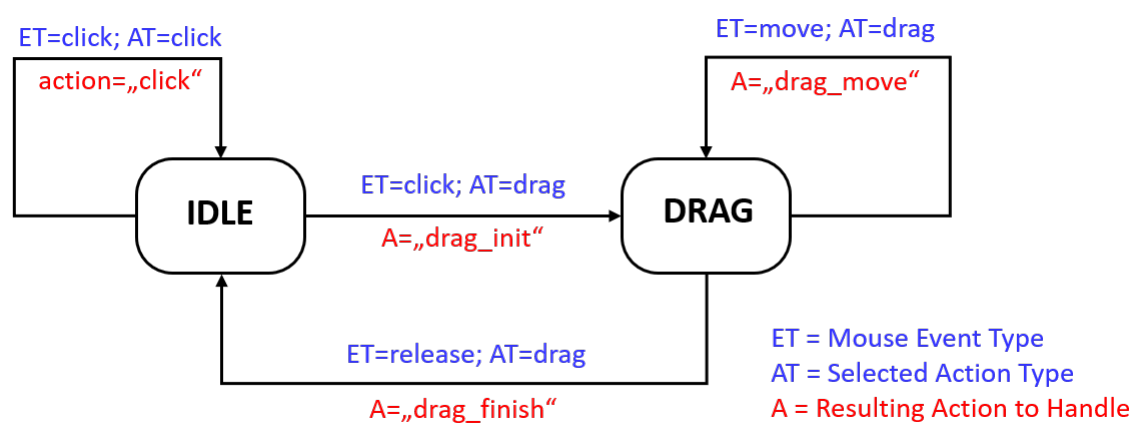}    
\end{center}
\caption{\label{fig:ClickDragEventAutomaton}
  State automaton realized by {\bf ivisit.special.ClickDragEventAutomaton} for handling mouse click and drag events.
  There are 2 states 'IDLE' and 'DRAG' (nodes) and 4 state transitions (edges). Conditions for transitions
  are displayed above each edge (event type ET and action type AT; blue), whereas the resulting action
  to be handled is displayed below each edge (action A; red). See text for details.
} 
\end{figure}

Creating interactive IVISIT/Matplotlib widgets using the Tkinter event handling as described in the previous section~\ref{sec:example4_InteractiveMatplotlibGUI}
is most flexible, but the creation of the state automatons can lead to somewhat confusing code.
To alleviate this, in many application cases it is {\bf much more convenient to use predefined event automatons} from {\tt ivisit.special.py} (see section~\ref{sec:architecture}, page~\pageref{txt:item:ivisist_special}).
For example, {\tt ivisit.special .ClickDragEventAutomaton} can be used to {\bf implement simple click and drag\&drop actions} (see Fig.~\ref{fig:ClickDragEventAutomaton}).
The purpose of an {\bf Event Automaton} is to handle the Tkinter GUI events (e.g., mouse clicks or mouse movements) and to convert them into
{\bf actions} that can be processed by an action-handler function. By this, the programmer of an IVISIT application can focus on actions, and needs not
to think of the underlying mouse events and states. Fig.~\ref{fig:ClickDragEventAutomaton} shows that the {\bf ClickDragEventAutomaton}
has two {\bf states}, IDLE and DRAG (where IDLE is initially active),
and can handle two different {\bf action types}, click-actions and drag-actions.
A {\bf click-action} only requires a single mouse-click-event in the IDLE state.
By contrast, a {\bf drag action} switches after the first mouse-click-event to the DRAG state, and then processes an arbitrary number of
mouse-move-events until a final mouse-release event will switch back to the IDLE state. For each incoming event and given action type
(both printed blue above edges in Fig.~\ref{fig:ClickDragEventAutomaton}), the automaton will produced a corresponding {\bf action} as output
(printed red below edges in Fig.~\ref{fig:ClickDragEventAutomaton}): For example, for action-type ``click'', a mouse-click will produce {\bf action=''click''}.
By contrast, for action-type ``drag'', the first mouse click will produce {\bf action=''drag\_init''}, each following mouse-move-event will
produce {\bf action=''drag\_move''}, and the final mouse-release-event will produce {\bf action=''drag\_finish''}. The actions along with
the mouse positions (in either Tkinter Canvas or Matplotlib coordinates) are then passed to the {\bf action handler} that must be implemented
as method of the main simulation class derived from {\tt IVisit\_Simulation}.

With such event automatons, it becomes simple and convenient to {\bf implement interactive IVISIT applications} in a few lines of code.
For example, the following Listing~\ref{lst:InteractiveClassifiers} shows the code of an IVISIT application for interactive classification of
data points using either a linear ``least squares'' classifier or a ``Kernel Multi-Layer-Perceptron'' (Kernel-MLP), where class-specific data points can be added, moved and deleted
similarly as in the previous example (cf., Listing~\ref{lst:InteractiveDataGeneration} and Fig.~\ref{fig:run_demo04_InteractiveDataGeneration} in section~\ref{sec:example4_InteractiveMatplotlibGUI}),
but the mouse-event-handling is done in a {\bf single action handler} function {\tt handle\_action(action,pos,pos\_init)} (see lines 58-76 in Listing~\ref{lst:InteractiveClassifiers}),
whereas the previous example required for much simpler actions three different functions (compare to lines 45--72 in Listing~\ref{lst:InteractiveDataGeneration}).

Before going into the details of Listing~\ref{lst:InteractiveClassifiers}, let us briefly review the two classification algorithms:
The {\bf linear classifier} computes a model (or discriminant) function
(e.g., see \cite{Bishop:2006}, sect.~4.1.3 or \cite{Knoblauch:LernSys:2024}, sect.~\ref{IAS:sec:LeastSquaresClassifier}, eq.~\ref{IAS:eq:LeastSquares_Diskriminanzfunktion_Zweiklassen})
\begin{align}
 y(\cv{x}) = \cv{w}\transp\cvsym{\phi}(\cv{x}) = \sum_{j=1}^Mw_j\phi_j(\cv{x}) \in\setR     \label{eq:linearclassifier_model_y}
\end{align}
to discriminate between two classes, classifying input data vector $\cv{x}\in\setR^D$ as class 1 if $y(\cv{x})\ge 0$ or class 2 if $y(\cv{x})<0$.
In general, the $\phi_1(\cv{x}),\ldots,\phi_M(\cv{x})$ are arbitrary basis (or feature) functions, and $\cvsym{\phi}(\cv{x})\in\setR^M$ is called the feature vector
corresponding to input $\cv{x}$. Here we define $\cvsym{\phi}(\cv{x}):=\begin{pmatrix}1 & x_1 & \cdots & x_D\end{pmatrix}\transp$ just as ``extended input vector'',
where $\cv{x}$ is extended by constant 1 or bias-component in the first component. The optimal {\bf ``least-squares'' weight vector} $\cv{w}$
minimizing the ``sum-of-squared-errors'' for a given training data set $\mathcal{D}:=\{(\cv{x}_n,t_n)|n=1,\ldots,N\}$ of $N$ input data vectors $\cv{x}_n\in\setR^D$
and corresponding class labels (or targets) $t_n\in\{+1,-1\}$ is then (e.g., \cite{Knoblauch:LernSys:2024}, eq.~\ref{eq:W_SSE2_linmod_klassifikation})
\begin{align}
  \cv{w} = \left(\cvsym{\Phi}\transp\cvsym{\Phi}+\lambda\mat{I}\right)^{-1}\cvsym{\Phi}\transp\mat{T}  \label{eq:linearclassifier_weights}
\end{align}
where $\cvsym{\Phi}:=\begin{pmatrix}\cvsym{\phi}_1 &\cdots& \cvsym{\phi}_N\end{pmatrix}\transp$ is the $N\times M$ design matrix having the feature vectors $\cvsym{\phi}_n:=\cvsym{\phi}(\cv{x}_n)\in\setR^M$
in its rows $n=1,\ldots,N$, $\mat{T}:=\begin{pmatrix}t_1 &\cdots& t_N\end{pmatrix}\transp\in\setR^N$ is the corresponding vector of class labels,
and $\lambda\ge 0$ is a regularization coefficient that, for $\lambda>0$, increases numeric stability and ensures invertibility.

The {\bf Kernel-MLP} corresponds to a three-layer neural network $\cv{x}\rightarrow\cv{z}\rightarrow\cv{y}$ with model function
(e.g., \cite{Knoblauch:LernSys:2024}, eq.~\ref{eq:MLP_Bsp3_Cybenko_final})
\begin{align}
  y(\cv{x})=\cv{w}\transp\cv{z}(\cv{x})     \quad\mbox{for}\quad  \cv{z}(\cv{x})=\cv{h^z}(\cvsym{\Phi}\cvsym{\phi}(\cv{x})   \label{eq:KernelMLP_model_y}
\end{align}
where $\cvsym{\phi}(\cv{x})$ and $\cvsym{\Phi}$ are feature vector and design matrix as in (\ref{eq:linearclassifier_model_y}) and (\ref{eq:linearclassifier_weights}),
$h^z(a)$ is the (in general non-linear) {\bf kernel activation function} of layer $\cv{z}$ (where $\cv{h^z}(\cv{a})$ applies $h^z$ to each component of vector $\cv{a}$).
Here we use either $h^z(a)=a$ (linear kernel), $h^z(a)=\tanh(\frac{a}{\sigma})$ (tangens hyperbolicus kernel), or $h^z(a)=1-\exp(-\frac{a^2}{2\sigma})$ (simple Gauss kernel),
where $\sigma>0$ is a scaling/variance factor determining the ``width'' of the kernel function. The weight vector $\cv{w}$ for the output is determined by
(e.g., see \cite{Knoblauch:LernSys:2024}, eq.~\ref{eq:MLP_Bsp3_Cybenko_final})
\begin{align}
   \cv{w}\transp&=\mat{T}\transp\cdot(\cv{h^z}(\mat{\Phi}\mat{\Phi}\transp+\lambda\mat{I}))^{-1}   \label{eq:KernelMLP_weights}
\end{align}
where $\mat{K}:=\mat{\Phi}\mat{\Phi}\transp$ is also referred to as the Gram matrix, and $\lambda>0$ is a regularization coefficient as in (\ref{eq:linearclassifier_weights}).
Both classification models are implemented in the following IVISIT simulation application of listing~\ref{lst:InteractiveClassifiers}:

%\lstinputlisting[language={Python},caption={demo05\_InteractiveClassifiers.py},label={lst:InteractiveClassifiers}]{python/demo05_InteractiveClassifiers.py}
\begin{lstlisting}[language={Python},caption={demo05\_InteractiveClassifiers.py},label={lst:InteractiveClassifiers}]
import numpy as np, ivisit as iv, ivisit.special as ivsp
from ivisit.matplotlib import getMatplotlibFigure, getMatplotlibImage

#@IVISIT:SIMULATION & demo05_InteractiveClassifiers1
#@IVISIT:SLIDER & log_lmbda & [200,1] & [-9,9,4,0.1] & log_lmbda & -1 & float &  -8   
#@IVISIT:CHECKBOX & Options    & [class border,grid,errors] & str_options     & 000   
#@IVISIT:RADIOBUTTON & Classifier & [LstSquares,KernelMLP] & str_classif & LstSquares
#@IVISIT:RADIOBUTTON & kernel  & [linear,tanh,gauss]       & str_kernel  & linear
#@IVISIT:SLIDER & kernel_sigma & [200,1] & [1,1000,4,1] & krn_sig  & -1 & float & 1
#@IVISIT:LISTSEL& Action& [10,3]& [Test,New,Delete,Move]& str_action& -1& string& New
#@IVISIT:RADIOBUTTON & New Data Label & [Class1,Class2] & str_newdata_label & Class1
#@IVISIT:DICTSLIDER & New Data Parameters & [200,20,-1,2,10] & par_NewData & 0
#@IVISIT:DICTSLIDERITEM & N          & [0,500, 3,1]    & N        & int   & 3
#@IVISIT:DICTSLIDERITEM & Sigma_11   & [0, 5, 3,0.1]   & Sigma_11 & float & 1.0 
#@IVISIT:DICTSLIDERITEM & Sigma_22   & [0, 5, 3,0.1]   & Sigma_22 & float & 1.0
#@IVISIT:DICTSLIDERITEM & Sigma_12   & [-5, 5, 3,0.1]  & Sigma_12 & float & 0.0

#@IVISIT:IMAGE    & Input Data & 1.0    & [0,255]   & im_data     & int
#@IVISIT:TEXT_OUT & Results    & [20,5] & just_left & str_results

# (I) define parameters to be controlled by IVisit
class SimParameters(iv.IVisit_Parameters):
    log_lmbda = -8           # regularization parameter lambda = 10^log_lmbda 
    str_options='000'        # select [class border,grid,errors] by '1' 
    str_classif='LstSquares' # selected classifier (LstSquares or KernelMLP)
    str_kernel ='linear'     # kernel function for hidden layer KernelMLP
    krn_sig  = 1.0           # scale/variance parameter of kernel functions
    str_action = 'Test'      # data action mode (Test data, New data, Move data)
    str_newdata_label='Class 1' # class label for new Gauss data
    par_NewData={'N':5,'Sig11':1.0,'Sig22':1.0,'Sig12':0} # num./cov. new data 
    x1min,x1max,x2min,x2max, dx1,dx2 = -8,8,-8,8, 0.1,0.1 # axis limits/grid constants
    
# (II) define simulation data to be displayed by IVisit
class SimData(iv.IVisit_Data):
    str_results = ''                  # string for text output
    im_data  = np.array([[0,0,0]])    # image of data vectors

# (III) main class for simulation 
class Sim(iv.IVisit_Simulation):
    def __init__(self,name_arg="demo05_InteractiveClassifiers.py"):
        iv.IVisit_Simulation.__init__(self,name_arg,SimParameters,SimData,flagLock=1)
        self.ax1=None         # reference to matplotlib axis
        self.data_event_autom=ivsp.ClickDragEventAutomaton(   # create automaton
            self, "str_action", "handle_action",              # for handling of
            {'Test':'click','New':'click','Delete':'click','Move':'drag'}, # actions
            flagMatplotlib=1) # involving mouse events on Matplotlib widgets
        self.init()           # already call init()
        
    def init(self):
        self.X=np.array([[2,1],[2,2],[2,-1],[3,-2],[-2,-1],[-1,-1]],'float') # data  
        self.T=np.array([1    ,1    ,1     ,-1    ,-1     ,-1]) # labels (-1=class 2)
        self.x=[0.0,0.0] # test data vector to classify (move in action mode 'Test') 
        
    def bind(self,parent,display): # parent=ivisit object (having display to bind to)
        self.data_event_autom.bind(display,"im_data","Input Data","ax1")

    def handle_action(self,action,pos,pos_init,pos_prev): # handle automaton actions
        p,pnd=SimParameters,SimParameters.par_NewData # short hands to parameters
        if p.str_action=='Test':   # update position of test vector x ?
            self.x=pos             # new position is clicked position
        if p.str_action=='New':    # insert new (random) data points with labels? 
            Sig = [[pnd['Sigma_11'],pnd['Sigma_12']], # covariance matrix of Gaussian
                   [pnd['Sigma_12'],pnd['Sigma_22']]] # to generate new data vectors
            X_new = np.random.multivariate_normal(pos,Sig,(pnd['N'])) # new data 
            self.X=np.concatenate((self.X,X_new)) # append new data vectors to data X
            t = 1 if p.str_newdata_label=='Class1' else -1      # add new data labels
            self.T=np.concatenate((self.T,t*np.ones(pnd['N']))) # to label/targets T
        elif p.str_action=='Delete' and len(self.X)>0:  # Delete (closest) data point?  
            self.idx_NN=ivsp.getNearestNeighbor(self.X,np.array(pos)) # get N.N. index 
            self.X=np.delete(self.X,self.idx_NN,axis=0) # delete nearest-neighbor data
            self.T=np.delete(self.T,self.idx_NN,axis=0) # delete corresponding label
        elif p.str_action=='Move' and len(self.X)>0: # move data point?
            if action=='drag_init': # first click of drag/move sequence then...
                self.idx_NN=ivsp.getNearestNeighbor(self.X,np.array(pos)) # ...get NN
            self.X[self.idx_NN,:]=pos # ... and update position of NN data point
                
    def step(self):
        # (i) train a classifier
        p,d=SimParameters,SimData       # short hand to simulation parameters and data
        lmbda=10**p.log_lmbda # regularization coefficient (for matrix invertibility)
        N,D=self.X.shape[0],self.X.shape[1]  # number and dim. of training data
        PHI = np.ones((N,D+1)) # allocate memory for design matrix (for extended data)
        PHI[:,1:]=self.X     # for design matrix extend X by 1-column (bias component)
        if p.str_classif=='LstSquares':  # least squares classifier?
            Pinv = np.linalg.inv(np.dot(PHI.T,PHI)+lmbda*np.eye(D+1)) # inverse matrix
            w    = np.dot(np.dot(Pinv,PHI.T),self.T)   # optimal least squares weights
            y_pred = lambda x: np.dot(w,[1,x[0],x[1]]) # simple linear predict. model
        elif p.str_classif=='KernelMLP': # Kernel multi-layer-perceptron?
            if p.str_kernel=='linear': hz=lambda a:a    # linear activation function?
            elif p.str_kernel=='tanh' :hz=lambda a:np.tanh(a/p.krn_sig)     # or tanh
            elif p.str_kernel=='gauss':hz=lambda a:1-np.exp(-a*a/p.krn_sig) # or Gauss
            K = hz(np.dot(PHI,PHI.T))  # Gram matrix for chosen kernel (act.) function
            w = np.dot(self.T.T,np.linalg.inv(K+lmbda*np.eye(N))) # weights Kernel-MLP
            z_pred = lambda x: hz(np.dot(PHI,[1,x[0],x[1]]))   # computes hidden layer
            y_pred = lambda x: np.dot(w,z_pred(x))             # prediction model

        # (ii) test classifier on training data and with new data point self.x
        y = y_pred(self.x)               # prediction/discr.value for new inpt self.x
        Y = np.array([y_pred(x) for x in self.X]) # predictions for trainings data
        clf_err = (Y>=0)!=(self.T>=0)    # array/indexes of classification errors 
        n_err = np.sum(clf_err)          # number classification errors train data

        # (iii) write results
        X1,X2=self.X[self.T>0],self.X[self.T<0]
        d.str_results="-------------------------------------\n"
        d.str_results+="number of training data N="+str(N)+"\n"
        d.str_results+="class1 N1="+str(X1.shape[0])+\
                       ", class2 N2="+str(X2.shape[0])+"\n"
        d.str_results+="lmbda="+str(lmbda)+"\n\n"
        d.str_results+="classification errors: n_err="+str(n_err)+\
                       ", p_err="+str(round(n_err/N,4))+"\n\n"
        d.str_results+="new input x="+str(np.round(self.x,4))+"   -->   y="+\
                       str(round(y,4))+", class="+str(1 if y>0 else 2)+"\n"
        
        # (iv) plot data
        fig = getMatplotlibFigure(figsize=(6,5)) # create figure
        ax1 = fig.add_subplot(111)               # create axis for data/contour plot
        ax1.scatter(X1[:,0],X1[:,1],c='r',marker='x',s=20) # plot data from class 1/2
        ax1.scatter(X2[:,0],X2[:,1],marker='o',s=20,facecolors='none',edgecolors='b')
        ax1.scatter([self.x[0]], [self.x[1]], marker='*',s=40, facecolors='k',
                    edgecolors='k') # plot new input vector
        if p.str_options[0]=='1':                # draw class borders/contour plot?
            x1g,x2g,yg = ivsp.computeGrid2D(y_pred,[p.x1min,p.x2max,p.dx1],
                             [p.x2min,p.x2max,p.dx2],f_mode='scalar')
            CS = ax1.contour(x1g, x2g, yg,[-2,0,2])  # contour plot
            ax1.clabel(CS, inline=True, fontsize=10) # define contour labels
        if p.str_options[1]=='1': ax1.grid(which='both') # draw a grid ?
        if p.str_options[2]=='1':
            ax1.scatter(self.X[clf_err,0],self.X[clf_err,1],marker='o',s=150,
                        facecolors='none', edgecolors='m')
        ax1.set_xlabel('x1'); ax1.set_ylabel('x2')   # labels on x/y-axis
        ax1.set_xlim((p.x1min,p.x1max))          # limits for axis
        ax1.set_ylim((p.x2min,p.x2max))          # limits for axis
        d.im_data = getMatplotlibImage(fig) # send matplotlib image to ivisit image
        self.ax1=ax1                     # save axis variable (for event automaton)
        return
               
# (IV) main program
iv.IVisit_main(sim=Sim(),str_app_name='demo05_InteractiveClassifiers')
\end{lstlisting}

The {\bf ClickDragEventAutomaton} named {\tt data\_event\_autom} is {\bf created} in the constructor of the main simulation class (lines 43--46), passing the following parameters: {\bf 1)} {\tt var\_action\_name='str\_\-action'} defines that the simulation parameter {\tt str\_action} (see lines 28 and 10) contains the currently active action as a string. {\bf 2)} {\tt handle\_action='handle\_action'} defines that function {\tt han\-dle\_action(.)} (lines 57--75) will be called to handle the actions corresponding to the mouse events (see Fig.~\ref{fig:ClickDragEventAutomaton}). {\bf 3)} {\tt dict\_action\_type=\{'Test':'click','New':'click','Delete':'cli\-ck','Move':'drag'\}} defines the action type for each possible action in {\tt str\_action}. Here actions {\tt 'Test'}, {\tt 'New'}, and {\tt 'Delete'} have action-type ``click'', whereas action {\tt 'Move'} has action-type ``drag'' (see Fig.~\ref{fig:ClickDragEventAutomaton}). {\bf 4)} {\tt flagMatplotlib=1} defines that click positions will be given as Matplotlib coordinates (instead of IVISIT canvas coordinates). {\bf 5)} The {\tt button\_no=1} (not given here, therefore defaulting to value 1) defines which mouse button to operate on (valid values are buttons 1=left, 2=middle, 3=right). Then the {\bf bindings} are done in function {\tt bind(.)} (line 55) calling
{\tt data\_event\_autom.bi\-nd(display,"im\_data","Input Data","ax1")} to bind mouse events on the IVISIT IMAGE widget ``Input Data'' (line 18) to the {\tt ClickDragEventAutomaton}. The further parameters mean that the IVISIT IMAGE is showing the simulation data array {\tt im\_data} (line 36) showing the Matplotlib figure axis {\tt ax1} (lines 42, 117, 135) on the TKinter frame {\tt display} .
Finally, the {\bf action handling} is defined
by function {\tt handle\_action(self,action,pos,pos\_init,pos\_prev)} (lines 57--75) passing the current {\tt action}, the current mouse position {\tt pos}, the initial mouse position  {\tt pos\_init}, and the previous mouse position {\tt pos\_prev}
(the latter two parameters being relevant only for action-type drag): {\bf 1) Action ``Test''} (lines 69--60) defines a new input vector
{\tt self.x} at the current mouse click position to be classified. {\bf 2) Action ``New''} (lines 61--67) creates {\tt N} new data vectors {\tt X\_new} as 2D-Gaussian random number (line 64), where the mean corresponds to the current mouse position {\tt pos}, and the number {\tt N} and the components of the covariance matrix {\tt Sig} (lines 62--63) are taken from the IVISIT DICTSLIDER {\tt New Data Parameters} (lines 12--16). The new data are labeled with +1 or -1 (line 66) depending on IVISIT RADIOBUTTON widget ``New Data Label'' (lines 11, 29), and finally concatenated with the old data (lines 64, 67). {\bf 3) Action ``Delete''} (lines 68-71) gets the nearest-neighbor data point w.r.t. the current mouse click position (line 69) and removes the data point together with its label (lines 70, 71).
% Figure fig_demo05_InteractiveClassifiers 
% created with the Ubuntu Bildschirmfoto-Tool and saved as png in ~/hs-albsig/02-research/KICAD/ivisit_techreport/figures/snapshots
% converted to pdf using img2pdf in the Windows operation system
\begin{figure}[hb]
%\nopagenumber
%\renewcommand{\baselinestretch}{1.0}
\hfill
\begin{center}
\includegraphics[width=\linewidth]{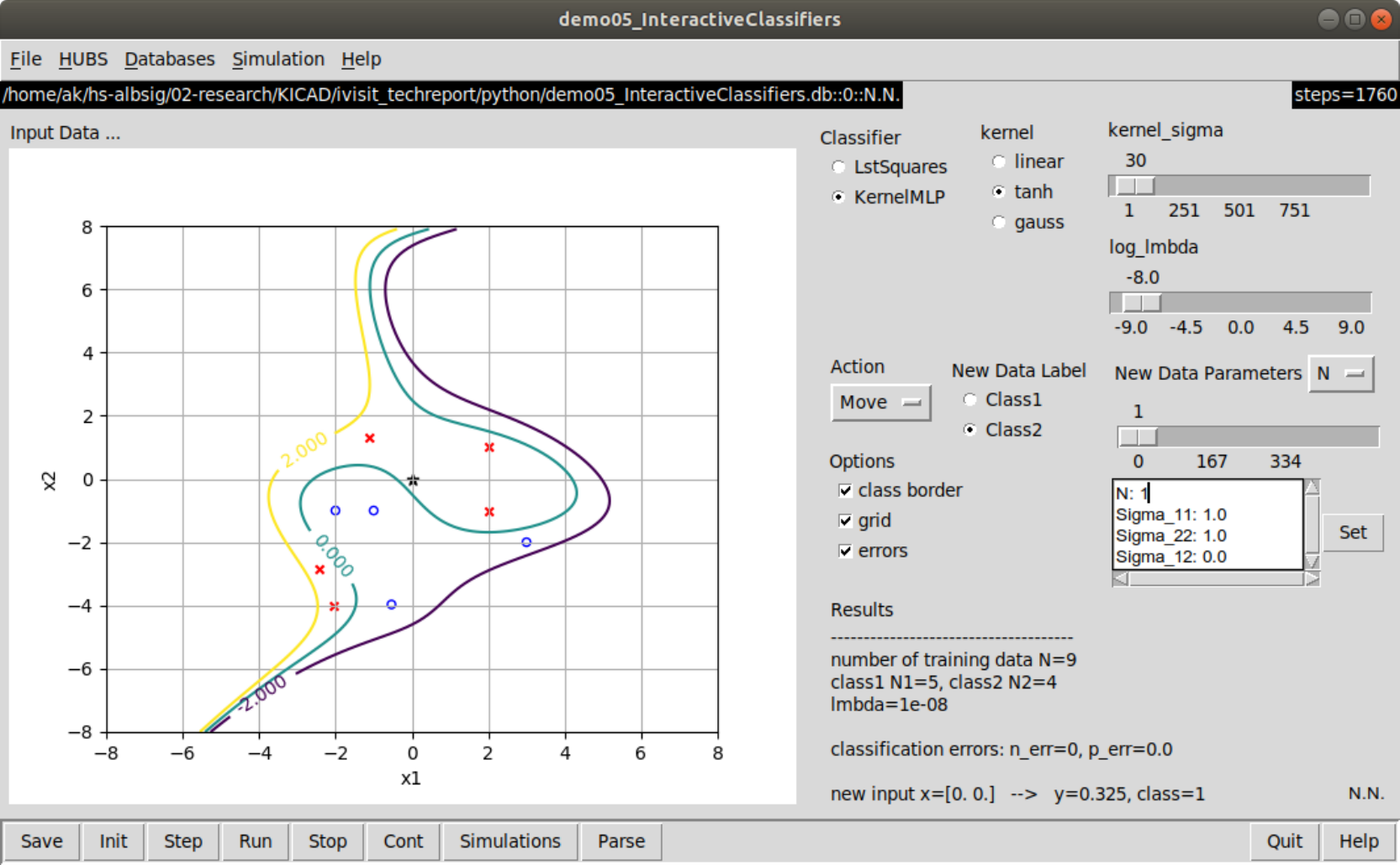}    
\end{center}
\caption{\label{fig:run_demo05_InteractiveClassifiers}
  Running the IVISIT application of Listing~\ref{lst:InteractiveClassifiers}
  for interactive classification of data points from two classes (red crosses vs. blue circles, corresponding to classes +1 vs. -1).
  Data from both classes can be added, deleted, or moved by clicking on the Matplotlib IMAGE widget ``Input Data''.
  Classes, classifiers and many further (hyper-) parameters can be modified interactively
  using the parameter widgets. See text for further details. 
} 
\end{figure}
{\bf 4) Drag action ``Move''} (lines 72--75) first checks if the passed {\tt action} is {\tt drag\_init} which means that the mouse button has been clicked (see Fig.~\ref{fig:ClickDragEventAutomaton}) and, if true, gets the index of the nearest-neighbor data point (line 74). After that, in any case ({\tt action} being either {\tt drag\_init}, {\tt drag\_move} or {\tt drag\_finish}; see Fig.~\ref{fig:ClickDragEventAutomaton}), the position of the nearest-neighbor data point will be updated to the current mouse position (line 75).

The remaining code defines the {\bf simulation step} function (lines 77--136) implementing the two classification models and displaying the data: Line 60 computes the {\bf regularization coefficient} $\lambda$ used in (\ref{eq:W_SSE2_linmod_klassifikation},\ref{eq:KernelMLP_weights}) from the logarithmic representation of IVISIT SLIDER widget "log\_lmbda'' (lines 5, 23). Lines 81--83 create the {\bf design matrix} $\cvsym{\Phi}$ or {\tt PHI} used in (\ref{eq:W_SSE2_linmod_klassifikation},\ref{eq:KernelMLP_weights}) by appending a column of 1s in front of the data matrix {\tt X}. In case of {\bf least-squares-classification} (line 84), lines 85-86 compute the
{\bf optimal weight vector} $\cv{w}$ according to (\ref{eq:linearclassifier_weights}), and line 87 defines the discriminant function $y(\cv{x})$ according to (\ref{eq:linearclassifier_model_y}). Similarly, in case of {\bf Kernel-MLP} (lines 88--95),
the Gram Matrix $\mat{K}$, output weights $\cv{w}$ and discriminant function $y(\cv{x})$ are computed according to (\ref{eq:KernelMLP_weights}) and (\ref{eq:KernelMLP_model_y}).
Then the classifier is {\bf tested on the new input} (line 98) {\bf and the training data} (lines 99--101)
by evaluating the discriminant function for each input vector (line 99)
and counting the number of classification errors (lines 100--101). Lines 103--113 {\bf write the results} to the IVISIT TEXT\_OUT widget {\tt Results} (lines 19, 35),
and lines 116--136 {\bf create a Matplotlib figure} showing the data points, class borders, and errors
in IVISIT IMAGE widget {\tt Input Data} (lines 18, 36),
depending on IVISIT CHECKBOX widget {\tt Options} (lines 6, 24). Finally, line 139 creates and starts the IVISIT main simulation object.

Fig.~\ref{fig:run_demo05_InteractiveClassifiers} shows a {\bf snapshot from running} Listing~\ref{lst:InteractiveClassifiers}, where the {\bf Kernel-MLP} with $\tanh$-Kernel has been selected. A few additional data points have been created and moved. The class border corresponds to the cyan contour line ``y=0.000'', indicating all data points where the discriminant function $y(\cv{x})$ has value 0. The plot also shows the contour lines $y=+2.000$
and $y=-2.000$ in yellow and violet. Note that all data points have been classified correctly. 

% *******************************************************************************************
\subsection{Example 6: Interactive training of a CNN with MNIST dataset using Keras}  \label{sec:example6_InteractiveTrainingKerasMNIST}
% *******************************************************************************************
As a final example, listing~\ref{lst:CNN_MNIST} presents a larger IVISIT application integrating most of the previous elements to create
a Convolutional Neural Network (CNN) for classification of handwritten digits that can be interactively trained
and tested using the MNIST data set and the Tensorflow/Keras machine learning library \cite{LeCun/Bottou/Bengio/Haffner:MNIST:1998,Deng:MNIST:2012,AbadiTensorflow:2016,CholletKeras:2015}:

%\lstinputlisting[language={Python},caption={demo06\_CNN\_MNIST.py},label={lst:CNN_MNIST}]{python/demo06_CNN_MNIST.py}
\begin{lstlisting}[language={Python},caption={demo06\_CNN\_MNIST.py},label={lst:CNN_MNIST}]
import numpy as np, ivisit as iv, ivisit.special as ivsp, ivisit.matplotlib as ivml
import cv2, scipy, tensorflow as tf
from tensorflow.keras.models import Sequential
from tensorflow.keras.layers import Conv2D, MaxPooling2D, Dense, Flatten
from tensorflow.keras.utils import to_categorical
from supy.computervision import getInt8Image

#@IVISIT:SIMULATION & demo06_CNN_MNIST1
#@IVISIT:DICTSLIDER & Network-Parameters & [200,20,-1,2,10] & par_Network & 0
#@IVISIT:DICTSLIDERITEM & num_filters & [1,50,2,1] & num_filters & int & 8
#@IVISIT:DICTSLIDERITEM & filter_size & [1,15,2,2] & filter_size & int & 3
#@IVISIT:DICTSLIDERITEM & pool_size   & [1,10,2,1] & pool_size   & int & 2
#@IVISIT:DICTSLIDER & Training-Parameters & [200,20,-1,2,10] & par_Training & 0
#@IVISIT:DICTSLIDERITEM  & batch_size & [1,60000, 2,1] & batch_size & int & 500 
#@IVISIT:DICTSLIDERITEM  & epochs     & [1,100,2,1]    & epochs     & int & 5 
#@IVISIT:DICTSLIDERITEM  & learn_rate & [1e-6,10,2,1e-6] & lrn_rt & float & 0.001
#@IVISIT:DICTSLIDERITEM  & beta_1     & [1e-6,1,2,1e-6]  & beta_1 & float & 0.9
#@IVISIT:DICTSLIDERITEM  & beta_2     & [1e-6,1,2,1e-6]  & beta_2 & float & 0.999
#@IVISIT:DICTSLIDER & Drawing-Parameters & [200,20,-1,2,10] & par_Drawing & 0
#@IVISIT:DICTSLIDERITEM  & linewidth & [1, 5, 2, 1]    & linewidth & int   & 1
#@IVISIT:DICTSLIDERITEM  & linecolor & [0, 255,2,1]    & linecolor & int   & 255
#@IVISIT:DICTSLIDERITEM  & shiftx    & [-15,15,2,1]    & shiftx    & int   & 0
#@IVISIT:DICTSLIDERITEM  & shifty    & [-15,15,2,1]    & shifty    & int   & 0
#@IVISIT:DICTSLIDERITEM  & scale     & [0.5,2.0,2,0.1] & scale     & float & 1.0
#@IVISIT:DICTSLIDERITEM  & rotate    & [-180,180,2,1]  & rotate    & float & 0
#@IVISIT:DICTSLIDERITEM  & p_noise   & [0,1,2,0.001]   & p_noise   & float & 0
#@IVISIT:DICTSLIDERITEM  & sig_noise & [0,100,2,0.1]   & sig_noise & float & 1
#@IVISIT:SLIDER & idx_image   & [150,1] & [0,69999,2,1] & idx_image   & -1 & int & 0 
#@IVISIT:SLIDER & scale_image & [150,1] & [1,20   ,2,1] & scale_image & -1 & int & 5 
#@IVISIT:SLIDER & filter_img_class &[150,1]&[-1,9,2,1]&filter_img_class&-1 & int &-1 
#@IVISIT:LISTSEL & Img-Source &[20,5]&[ALL,TRAIN,TEST]&str_image_source&-1&string&ALL
#@IVISIT:RADIOBUTTON & Input Source & [image,new_input] & str_input_source & image
#@IVISIT:CHECKBOX    & Options      & [plot eval,grid,legend]& str_options & 011   
#@IVISIT:BUTTON  & bt_build        & [<,BUILD NETWORK]   & bt_build
#@IVISIT:BUTTON  & bt_compile      & [<,COMPILE NETWORK] & bt_compile
#@IVISIT:BUTTON  & bt_train        & [<,TRAIN NETWORK]   & bt_train
#@IVISIT:BUTTON  & bt_newimg_erase & [<,ERASE_NEWIMAGE]  & bt_newimg_erase
#@IVISIT:BUTTON  & bt_newimg_copy  & [<,COPY_NEWIMAGE]   & bt_newimg_copy
#@IVISIT:BUTTON  & bt_newimg_add   & [<,ADD_NEWIMAGE]    & bt_newimg_add
#@IVISIT:IMAGE    & MNIST Image     & 1.0    & [0,255]   & im_image      & int
#@IVISIT:IMAGE    & New Input       & 1.0    & [0,255]   & im_inp_new_sc & int
#@IVISIT:IMAGE    & Network Outputs & 1.0    & [0,255]   & im_outputs    & int
#@IVISIT:IMAGE    & Evaluation Scores & 1.0  & [0,255]   & im_eval       & int
#@IVISIT:TEXT_OUT & Results         & [20,7] & just_left & txt_results 

# (I) define parameter variables to be controlled by IVisit
class SimParameters(iv.IVisit_Parameters):
    idx_image    = 0       # index of image to be displayed
    scale_image  = 5       # scale factor for increasing size of image
    filter_img_class = -1  # filter for one particular class (-1 means all classes)
    par_Network  = {'num_filters':8, 'filter_size':3, 'pool_size':2} # compiling CNN
    par_Training = {'batch_size':500, 'epochs':5, 'lrn_rt':0.001,   # parameters for
                    'beta_1':0.9,'beta_2':0.999}    # training (ADAM + policy agent)
    par_Drawing  = {'linewidth':1,'linecolor':255,'shiftx':0,'shifty':0, # for drawing
                    'scale':1.0,'rotate':0.0,'p_noise':0,'sig_noise':1.}   # new image 
    bt_build    = '0' # if '1' (pressed) then build neural network with network param.
    bt_compile  = '0' # if '1' (button pressed) compile neural network 
    bt_train    = '0' # if '1' (button pressed) then train neural network 
    bt_evaluate = '0' # if '1' (button pressed) evaluate neural network (test data)
    bt_newimg_erase = '0'    # if '1' then erase new image
    bt_newimg_copy  = '0'    # if '1' then copy image to new_image
    bt_newimg_add   = '0'    # if '1' then add image to new_image
    str_image_source='ALL'   # where to select image from
    str_input_source='image'
    str_options='00'         # select [plot eval,grid] by '1' 
    dx,pad,szy=10,10,50 # parameters of class probability bars (width,spacing,height)  
    
# (II) define simulation data to be displayed by IVisit 
class SimData(iv.IVisit_Data):
    txt_results   = ''                  # text output showing selected results
    im_image      = np.array([[0,0,0]]) # Input Image
    im_input_new  = np.zeros((28,28),'uint8') # New Input Image (MNSIT size 28x28)
    im_inp_new_sc = np.array([[0,0,0]]) # New Inp. Img (scaled for display/drawing)
    im_outputs    = np.array([[0,0,0]]) # image of output distribution (class prob.)
    im_eval       = np.array([[0,0,0]]) # image of evaluation scores (Matplotlib) 

# (III) main class for simulation  
class Sim(iv.IVisit_Simulation):
    def __init__(self,name_arg="ivisit_cnn_mnist.py"):
        iv.IVisit_Simulation.__init__(self,name_arg,SimParameters,SimData)
        self.data_event_autom=ivsp.ClickDragEventAutomaton(   # create automaton
            self, None, "handle_action", "drag")    # for handling of drag actions
        self.init()

    def init(self):
        self.model = None     # Keras neural network model
        self.flagCompiled=0   # if >0 then model has already been compiled
        self.flagTrained=0    # if >0 then model has already been trained
        self.acc    ,self.loss    =[],[] # initialize training accuracy/loss lists
        self.val_acc,self.val_loss=[],[] # initialize validation accuracy and loss
        
    def bind(self,parent,display): # parent=ivisit object (having display to bind to) 
        self.data_event_autom.bind(display,"im_input_new","New Input") # bind to IMAGE

    def handle_action(self,action,pos,pos_init,pos_prev): # handle (drag) actions
        p,d=SimParameters,SimData # short hands to simulation parameters and data 
        cv2.line(d.im_input_new,pos_prev,pos,      # draw line to new input image
                 p.par_Drawing['linecolor'],int(p.par_Drawing['linewidth']))

    def step(self):
        # (i) display selected image from MNIST data set 
        p,d=SimParameters,SimData       # short hand to simulation parameters and data
        idx_list=dict_img_idx[p.str_image_source+'_'+str(p.filter_img_class)] #idxlist
        if not self.parent.display is None: # adjust range of slider to selected class
           self.parent.display.getWidget("idx_image").reconfigure(max=len(idx_list)-1)               
        idx_img = idx_list[p.idx_image%len(idx_list)] # get index of selected image
        if idx_img>=60000:    # indexes >=60000 are test images 
            im=test_images[(idx_img-60000)%10000]     # get test image
            im_cls=test_labels[(idx_img-60000)%10000] # corresponding label
        else:                 # indexes <60000 are training images
            im=train_images[idx_img%60000]            # get train image
            im_cls=train_labels[idx_img%60000]        # corresponding label
        d.im_image=getInt8Image(im.reshape((28,28)))
        newsize=(d.im_image.shape[1]*p.scale_image,d.im_image.shape[0]*p.scale_image)
        if p.scale_image>1:   # scale image to new size? 
            d.im_image=cv2.resize(d.im_image,newsize,interpolation=cv2.INTER_NEAREST)

        # (ii) draw new input image
        # (ii.1) check action buttons for erasing, copying or adding MNIST images
        if p.bt_newimg_erase=='1':d.im_input_new[:,:]=0    # erase new input image
        if p.bt_newimg_copy=='1':d.im_input_new[:,:]=getInt8Image(im.reshape((28,28)))
        if p.bt_newimg_add=='1':d.im_input_new[:,:]=np.minimum(d.im_input_new+\
                                                getInt8Image(im.reshape((28,28))),255)
        # (ii.2) do further transformations: rotation, scaling, shifting, noise 
        im_inp_new=np.array(d.im_input_new) # get copy for further transformations
        if p.par_Drawing['rotate']!=0:      # do rotation?
            im_inp_new=scipy.ndimage.rotate(im_inp_new,p.par_Drawing['rotate'],\
                                  reshape=False,mode='constant',cval=0,prefilter=True)
        if p.par_Drawing['scale']!=1.0:     # do scaling?
            new_sz=(round(im_inp_new.shape[1]*p.par_Drawing['scale']),\
                    round(im_inp_new.shape[0]*p.par_Drawing['scale'])) # scaled size
            im_inp_new=cv2.resize(self.im_input_new,new_sz,            # resize
                                  interpolation=cv2.INTER_NEAREST)     # input image
            if p.par_Drawing['scale']>1.0:  # is scaling an upsampling (larger image?)
                pad=(new_sz[0]-28)//2                  # then take central part of the 
                im_inp_new=im_inp_new[pad:(pad+28),pad:(pad+28)]      # enlarged image
            else:                           # else is scaling downsampling
                pad=(28-new_sz[0])//2                  # then fill up surrounding
                im_empty=np.zeros((28,28),'uint8')     # image with zeros
                im_empty[pad:(pad+new_sz[1]),pad:(pad+new_sz[0])]=im_inp_new # copy
                im_inp_new=im_empty         # scaled image is now in the center part
        if p.par_Drawing['shiftx']!=0:      # do shifting in x-direction?
            im_inp_new=np.roll(im_inp_new,p.par_Drawing['shiftx'],1)
        if p.par_Drawing['shifty']!=0:      # do shifting in y-direction?
            im_inp_new=np.roll(im_inp_new,p.par_Drawing['shifty'],0)
        if p.par_Drawing['p_noise']>0:      # add noise?
            mask_noise=np.zeros((28,28),'uint8') # init noise mask with 0s and select 
            mask_noise[np.random.rand(28,28)<p.par_Drawing['p_noise']]=1 # rnd pixels
            rnd_val=np.random.normal(p.par_Drawing['linecolor'], # get Gaussian random   
                                  p.par_Drawing['sig_noise'],(28,28)) # noise values
            im_new=np.maximum(np.minimum(np.round(im_inp_new+        # set mask pixels
                            np.multiply(mask_noise,rnd_val)),255),0) # to noise values
            im_inp_new=np.array(im_new,'uint8') # cast new input image as uint8 array
        d.im_inp_new_sc=cv2.resize(im_inp_new,newsize,      # scale new input image to 
                                   interpolation=cv2.INTER_NEAREST)  # IMG widget size
        
        # (iii) (re-) build, compile, and train neural network using Keras/Tensorflow?
        if p.bt_build=='1':           # (iii.1) (re-) build neural network?
            self.model = Sequential([ # create a sequential layered neural network            
                Conv2D(p.par_Network['num_filters'], p.par_Network['filter_size'],
                       input_shape=(28, 28, 1), name='layers_conv'),  # CNN layer
                MaxPooling2D(pool_size=p.par_Network['pool_size'],
                             name='layers_maxpool'),                  # Max-Pool layer
                Flatten(name='layers_flatten'),                     # Flatten 2D to 1D
                Dense(10, activation='softmax',name='layers_dense')]) # dense layer
            self.flagCompile=0        # freshly build network is not yet compiled..
            self.flagTrain=0          # ... and not yet trained
        if p.bt_compile=='1' and not self.model is None: # (iii.2) (re-) compile NN?
            self.optimizer=tf.keras.optimizers.Adam(
                learning_rate=p.par_Training['lrn_rt'],
                beta_1=p.par_Training['beta_1'],
                beta_2=p.par_Training['beta_2'])         # construct ADAM optimizer
            self.model.compile(
                optimizer=self.optimizer,
                loss='categorical_crossentropy',
                metrics=['accuracy'])                    # compile neral network model
            self.flagCompiled=1       # set flag to indicate that network was compiled
        if p.bt_train=='1' and self.flagCompiled>0:  # (iii.3) (re)train neur.network?
            fun_scheduler = lambda epoch,lr:p.par_Training['lrn_rt'] # scheduler for
                                      # updating of learning rate (here only constant)
            res=self.model.fit(                      # do training of neural network
                train_images,                        # input images for training
                to_categorical(train_labels),  # correspond. class labels (1-hot-code)
                batch_size=p.par_Training['batch_size'],  # minibatch size
                epochs=p.par_Training['epochs'],          # number of training epochs
                validation_data=(test_images, to_categorical(test_labels)), # val.data
                callbacks=[tf.keras.callbacks.LearningRateScheduler(fun_scheduler)])
                                       # callbacks will be called at end of each epoch 
            self.flagTrain=1           # set flag to indicate that network was trained 
            print("res=",res.history.keys())   # output some results on command window
            self.acc     +=res.history['accuracy']  # save train-data accuracy in list
            self.loss    +=res.history['loss']      # save train-data loss in list
            self.val_acc +=res.history['val_accuracy'] # save validation accuracy 
            self.val_loss+=res.history['val_loss']     # save validation loss in list

        # (iv) classify current image (or new input image)
        if p.str_input_source=='new_input':im=(im_inp_new/255)-0.5 # scale as MNIST im 
        if self.flagCompiled>0: y = self.model.predict(im.reshape((1,28,28,1)))[0]
                    # if compiled then predict class probabilities of current image im
        else: y = np.zeros(10)                 # else use dummy zero values as outputs
        y_hat=np.argmax(y)       # class decisions corresponds to largest output value
        szx=10*(p.dx+p.pad)+p.pad     # width of im_outputs for class probability bars
        d.im_outputs=255*np.ones((p.szy,szx,3),'uint8')     # allocate white RGB image 
        if y_hat==im_cls: col=(0,255,0)    # color green if predicted class is correct
        else: col=(255,0,0)                # color red if class is wrong
        for c in range(10):cv2.rectangle(d.im_outputs,(p.pad+c*(p.dx+p.pad),\
                         p.szy-p.pad),((c+1)*(p.dx+p.pad),p.pad+int((1-y[c])\
                         *(p.szy-2*p.pad))),col,3)  # draw bar for class c

        # (v) text output of results
        if len(self.acc)>0: acc,loss,val_acc,val_loss \
           = self.acc[-1],self.loss[-1],self.val_acc[-1],self.val_loss[-1] # acc/loss
        else: acc,loss,val_acc,val_loss=-1,-1,-1,-1       # or init with dummy values
        d.txt_results='trained '+str(len(self.acc))+' epochs:\n'
        d.txt_results+='train data: acc='+str(round(acc,4))\
                                         +'; loss='+str(round(loss,4))+'\n'
        d.txt_results+='validation data: acc='+str(round(val_acc,4))\
                                              +'; loss='+str(round(val_loss,4))+'\n'
        d.txt_results+='\ncurrent input:\n'
        d.txt_results+='idx_img='+str(idx_img)+'; predicted class y_hat='\
                                 +str(y_hat)+'; target class t='+str(im_cls)+'\n'
            
        # (vi) plot evaluation scores (using Matplotlib)
        if p.str_options[0]=='1':           # do plotting?
            fig = ivml.getMatplotlibFigure(figsize=(5,4)) # create figure
            ax1 = fig.add_subplot(111)      # create axis for evaluation scores
            epochs=range(1,len(self.acc)+1) # list of training epochs
            ax1.semilogy(epochs,1-np.array(self.acc),'r-',label='err_train') # errors
            ax1.semilogy(epochs,1-np.array(self.val_acc),'r--',label='err_test')
            ax1.semilogy(epochs,self.loss,'b-',label='loss_train') # plot train loss
            ax1.semilogy(epochs,self.val_loss,'b--',label='loss_test') # test loss
            if p.str_options[1]=='1': ax1.grid(which='both')          # draw a grid ?
            if p.str_options[2]=='1': ax1.legend(loc='upper right')   # draw legends?
            ax1.set_xlabel('# epochs'); ax1.set_ylabel('evaluation scores')  # labels 
            d.im_eval = ivml.getMatplotlibImage(fig)      # send plot to ivisit image

# (IV) main program
# (IV.1) load, normalize and reshape MNIST data 
train_data, test_data = tf.keras.datasets.mnist.load_data(path='mnist.npz')
train_images,train_labels = train_data[0],train_data[1]   # extract training data
test_images ,test_labels  = test_data[0] ,test_data[1]    # extract testing data
train_images = np.expand_dims( (train_images/255)-0.5, axis=3) # add feature map dim
test_images  = np.expand_dims( (test_images /255)-0.5, axis=3) # add feature map dim

# (IV.2) create indexes (for easy selection of images); indexes >=60000 are test img.
dict_img_idx={}
dict_img_idx['ALL_-1'  ]=list(range(70000))  # class "-1" refers to all classes (0-9)
dict_img_idx['TRAIN_-1']=list(range(60000))  # training data of all classes
dict_img_idx['TEST_-1' ]=[i+60000 for i in range(10000)] # test data of all classes
for c in range(10):                          # iterate over digit classes 0-9
    idx_c_train = [i for i in range(60000) if train_labels[i]==c] # train img class c
    idx_c_test  = [i+60000 for i in range(10000) if test_labels [i]==c] # idx test im  
    dict_img_idx['ALL_'  +str(c)]=idx_c_train+idx_c_test  # indexes of all images
    dict_img_idx['TRAIN_'+str(c)]=idx_c_train # indexes only of training images
    dict_img_idx['TEST_' +str(c)]=idx_c_test  # indexes only of testing images

# (IV.3) start IVISIT simulation application
iv.IVisit_main(sim=Sim(),str_app_name='demo06_CNN_MNIST')
\end{lstlisting}

The general structure of listing~\ref{lst:CNN_MNIST} is the same as in the previous examples: Lines~1--6 import the {\bf necessary libraries} .
This includes Tensorflow/Keras
for the Machine Learning models \cite{AbadiTensorflow:2016,CholletKeras:2015} and OpenCV for some computer vision methods like rotating
and resizing images or drawing lines and rectangles \cite{Bradski:OpenCV:2000}, which you may have to install (for example,
using {\tt pip install tensorflow, opencv-python}; see section~\ref{sec:examples}, page~\pageref{sec:examples}). Lines~8--4 define various IVISIT GUI widgets for parameter manipulation and displaying results. Lines~47--66 are the simulation parameters, and lines 69--75 the simulation data. Lines 78--235 is the simulation class, and lines 237--259 is the main program, loading, normalizing, reshaping, and
indexing the MNIST data set, and starting the main simulation. The {\bf indexing} (lines 246--255) is useful to select subsets of the MNIST data bases on the IVISIT LIST-SELECTION widget {\tt Img-Source} (ALL,TRAIN, or TEST, see line 31) and the class label (see IVISIT SLIDER widget {\tt filter\_img\_class}, line 30). To {\bf draw and test new input images}, line~81 in class {\tt Sim} defines again a {\tt ClickDragEventAutomaton}
(see lines 81, 82; cf., section~\ref{sec:example5_InteractiveGUI_IVISIT_eventautomatons}, Fig.~\ref{fig:ClickDragEventAutomaton}), and binds it to the IVISIT IMAGE widget {\tt New Input} (see lines 93, 41) for handling mouse drag events for drawing lines (see lines 95--98). Lines 100--235 define a {\bf simulation step}: Lines 101--116 copy the selected MNIST image into data array {\tt d.im\_image} and rescale it to the desired size. Lines 118--155 draw the new input image into data array {\tt d.im\_inp\_new\_sc} allowing for various operations like erasing, copying or adding the MNIST image, rotating, scaling, shifting, and addition of noise. Lines 157--194 define, compile, and train the CNN. It consists of a CNN layer, a max-pool layer, and (after flattening) a dense layer (lines 159--165). The network is compiled (lines 168--177) and trained (lines 178--194) after the corresponding IVISIT BUTTONS have been pressed (see lines 34--36). Lines 1996--208 test the network using as input either the selected MNIST image or the new input image. Lines 210--221 define the text output, and lines 223--235 do a Matplotlib plot of the train and testing evaluation scores.  

% Figure fig_demo06_CNN_MNIST 
% created with the Ubuntu Bildschirmfoto-Tool and saved as png in ~/hs-albsig/02-research/KICAD/ivisit_techreport/figures/snapshots
% converted to pdf using img2pdf in the Windows operation system
\begin{figure}[th]
%\nopagenumber
%\renewcommand{\baselinestretch}{1.0}
\hfill
\begin{center}
\includegraphics[width=\linewidth]{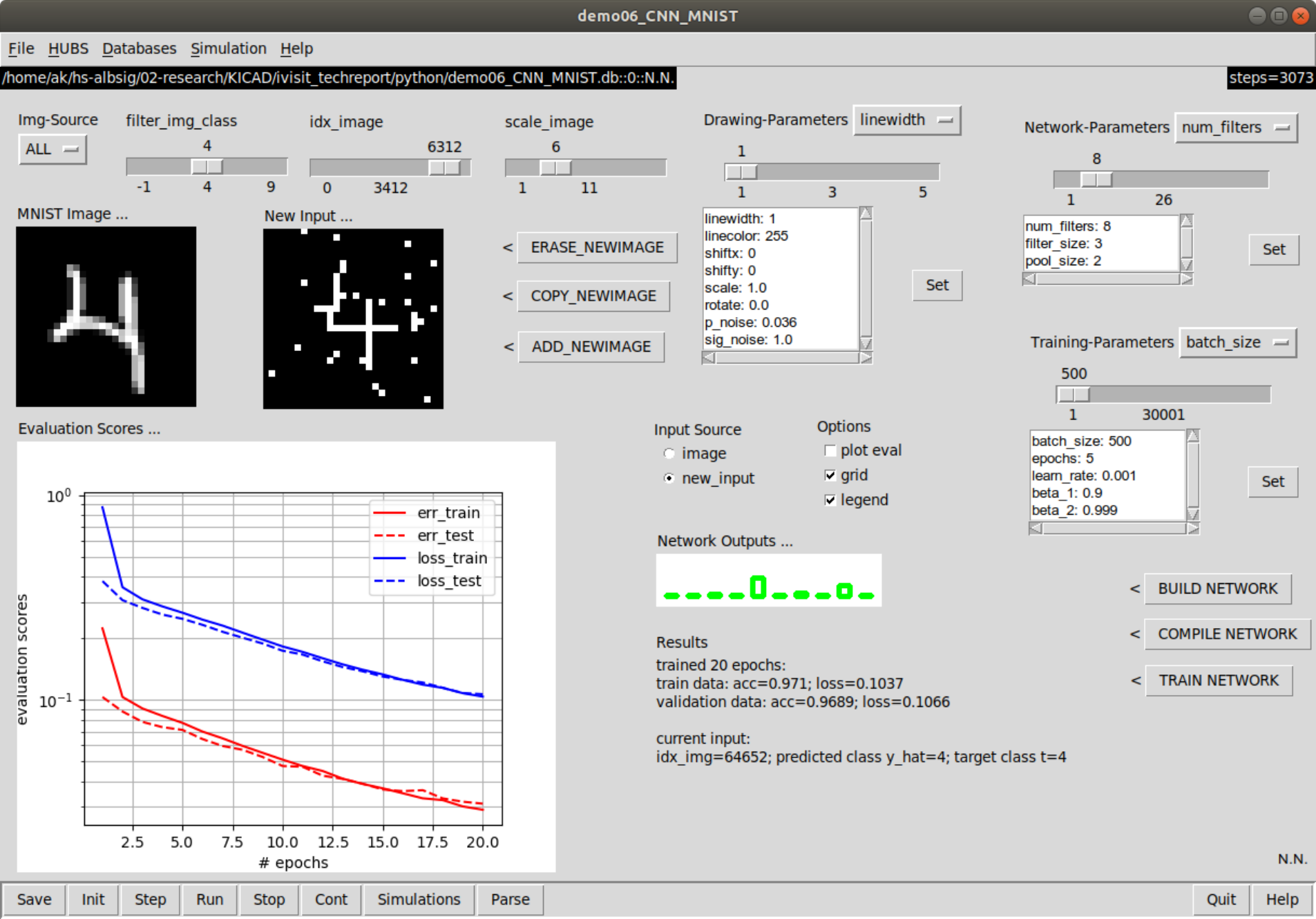}    
\end{center}
\caption{\label{fig:run_demo06_CNN_MNIST}
  Running the IVISIT application of Listing~\ref{lst:CNN_MNIST}
  for interactive training and evaluation of a Convolutional Neural Network (CNN) with the MNIST data set
  of handwritten digits. The network can also be tested by drawing digits into
  the IVISIT IMAGE widget ``New Input`` (after selecting ``new\_input'' from RADIOBUTTON widget ``Input Source``).
  Here, after 20 epochs of training (see IMAGE widget ``Evaluation Scores'') a ``4''  has been drawn
  by hand (see IMAGE widget ``New Input''), some noise has been added (see DICTSLIDER widget ``Drawing-Parameters''), and 
  the new input is still correctly classified as digit ``4'' (See IMAGE widget ``Network Outputs'' and TEXT widget ``Results'').
  See text for details. 
} 
\end{figure}

Figure~\ref{fig:run_demo06_CNN_MNIST} shows a snapshot from running the IVISIT simulation application: From the IVISIT DICTSLIDER {\tt Network-Parameters} (upper right) it can be seen that a CNN network with 8 feature maps ({\tt num\_filters}), filter kernel size $3\times 3$ ({\tt filter\_size}) and pooling size $2 \times 2$ ({\tt pool\_size}) has been created, and trained with minibatch size 500, for 5 epochs per click on button {\tt TRAIN NETWORK}, using the ADAM optimizer \cite{Kingma/Ba:ADAM:2015} with learning rate $0.001$ and parameters $\beta_1=0.9$, $\beta_2=0.999$ (see IVISIT DICTSLIDER {\tt Training-Parameters} (middle right)). The network has been build, compiled, and trained by clicking on the corresponding buttons (lower right). The IVISIT IMAGE widget {\tt Evaluation Scores} shows the training/test errors and the corresponding loss values as functions of training epochs (lower left). After 20 epochs the CNN-network achieves test error of about 3 percent. With the IVISIT CHECKBOX {\tt Options} you can choose if the evaluation plot is updated each simulation step (unchecking this will speed-up the simulation considerably, as drawing Matplotlib figures is time-consuming), or if a grid or the legends are displayed (middle panel). 
The panels on the upper left define selection of the MNIST image and creation of a new input image: Currently, MNIST images are selected from {\tt ALL} the available data (i.e., from TEST and TRAINING data; see IVISIT LIST-SELECTION {\tt Img-Source}; upper left). These data are filtered for images from class 4 (IVISIT SLIDER {\tt filter\_img\_class}; upper left), where the class-4 image with index 6312 is currently selected (IVISIT SLIDER {\tt idx\_image}; upper middle). The image sizes are scaled by factor 6 (IVISIT SLIDER {\tt scale\_image}; upper middle). The IVISIT IMAGE {\tt MNIST image} displays the selected class-4 image from the MNIST data set (mid left). Next to this the IVISIT IMAGE {\tt New Input} displays a hand-drawn image of digit ``4'' where some noisy pixels have been added. The new input can be created by hand-drawing (click-and-drag) into the IMAGE widget
(where click-and-drag events are handled by the {\tt ClickDragEventAutomaton}; see listing~\ref{lst:CNN_MNIST}, lines 81,82,93,95--98),
by using the IVISIT BUTTONS {\tt ERASE\_NEWIMAGE}, {\tt COPY\_NEWIMAGE}, {\tt ADD\_NEWIMAGE}, and by selecting parameters from IVISIT DICTSLIDER {\tt Drawing-Parameters} (middle upper).
From the latter you can select {\tt linewidth} and {\tt linecolor} for drawing by hand, {\tt shiftx} and {\tt shifty} to shift the image horizontally and vertically, {\tt scale} and {\tt rotate} to scale the image size and rotate it by an angle, and {\tt p\_noise} and {\tt sig\_noise} to define fraction of noise pixels and the standard deviation of additive noisy pixel values (drawn from a Gaussian with zero mean). The test input source for the CNN network can be selected with the IVISIT RADIOBUTTON widget {\tt Input Source} (middle panel; either {\tt image} or {\tt new\_input}). For the test input the IVISIT IMAGE widget {\tt Network Outputs} (lower middle) displays the activity of the 10 softmax output neurons
(representing the class probability distribution for the 10 digits 0-9). The largest bar corresponds to the class decision 4, and the green color indicates that the class decision is correct (color red means wrong). Finally, an overview of all results are displayed in the IVISIT TEXT widget {\tt Results} (middle lower panel).   

% *******************************************************************************************
% *******************************************************************************************
\section{Conlusions}    \label{sec:conclusions}  % for the original ideas see A.Knoblauch, AB4/p294 from 25/2/2023
% *******************************************************************************************
% *******************************************************************************************

The previous sections have given an overview and documentation of the IVISIT Interactive Visual Simulation Tool framework, along with a number of best-practice examples of writing IVISIT simulation applications. The examples demonstrate that, with only a few lines of codes, interesting interactive GUI applications can rapidly be created, and employed to simulate physical systems or test Machine Learning models and optimize hyperparameters. 
At Albstadt-Sigmaringen-University we have successfully used IVISIT applications similar to the previous examples of section~\ref{sec:examples} for practical student training
accompanying introductory courses on Machine Learning, as well as in various research and development projects, for example, for rapid creation of specialized labeling tools for image annotation,
to do psychophysical experiments on visual perception, simulating biological neural networks, 
and to conduct computational experiments and optimization with artificial neural networks and machine learning models. IVISIT is free software published under the MIT license.

\appendix

% *******************************************************************************************
% *******************************************************************************************
\section{UML class diagrams and SQL tables}    \label{app:UML_class_diagrams}   
% *******************************************************************************************
% *******************************************************************************************
Figure~\ref{fig:design_gensim} shows the original class design from the first IVISIT version in 2017
(the original name of IVISIT was GENSIM for ``generic simulation framework'',
and changed to IVISIT in 2019, because there existed already another software named ``GENSIM''). 
Although the classes have been modified, renamed and significantly extended since then,
the raw structure is still the same. There are {\bf three types of GUI widgets}:

% Figure figXX_design_gensim (UML design IVISIT)
% created with dia ~/projects/experimental/IVisit/design/design_gensim.dia
% exported as pdf
\begin{figure}[tp]
%\nopagenumber
%\renewcommand{\baselinestretch}{1.0}
\hfill
\begin{center}
\includegraphics[width=\linewidth]{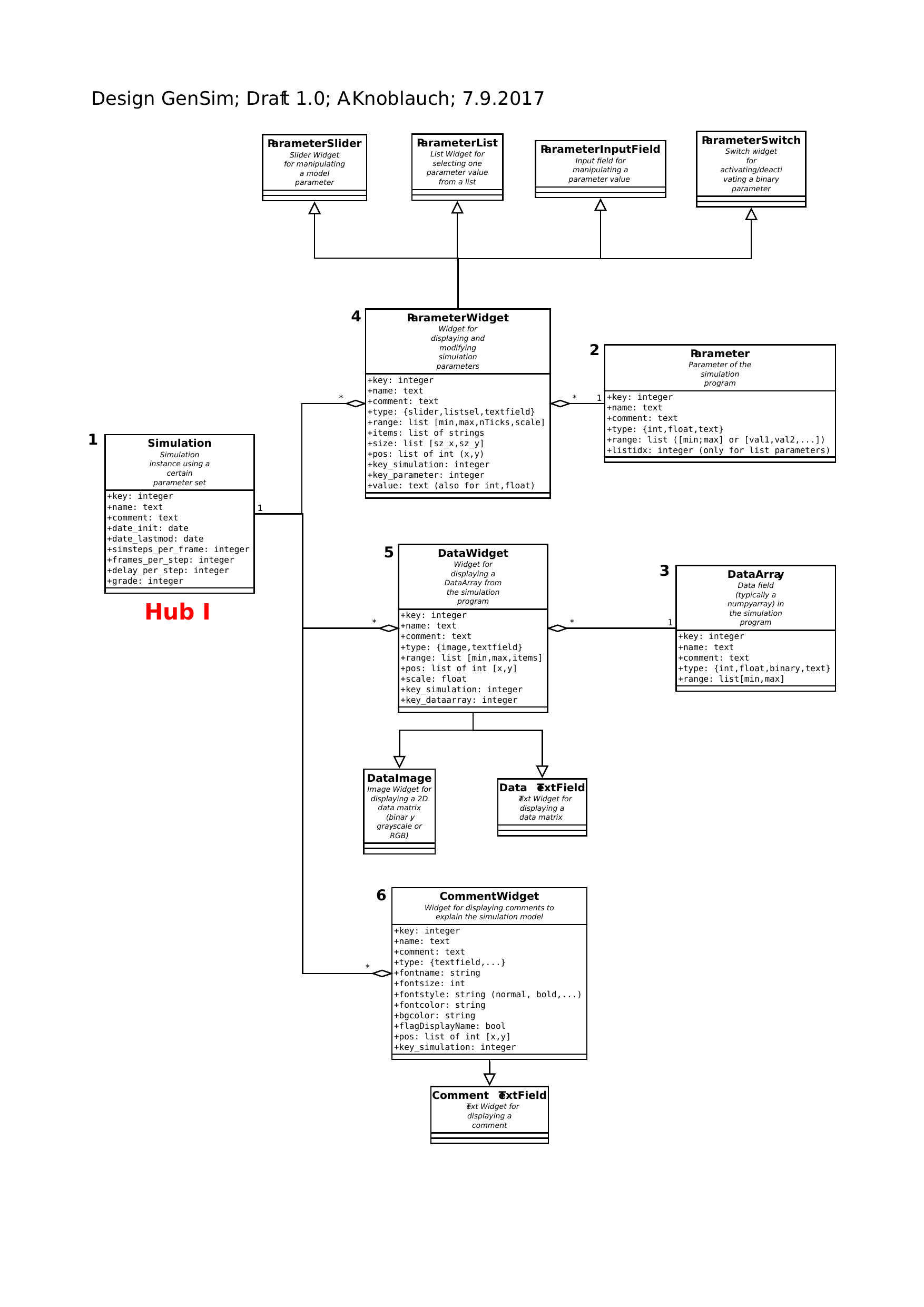}    
\end{center}
\caption{\label{fig:design_gensim}
  UML class diagram of the main IVISIT classes (from the 2017 original IVISIT/GENSIM version; see text for details). 
} 
\end{figure}

\begin{enumerate}\tightitems
\item {\bf Parameter Widgets} interface with Python variables used as simulation parameters (cast as class {\tt Parameter}).
  The subclasses ParameterSlider, ParameterList, ParameterInputField and ParameterSwitch displayed in Fig.~\ref{fig:design_gensim}
  correspond to subclasses derived from base class {\tt IVisitWidget} in module {\tt ivisit.widgets}
  (see section~\ref{sec:architecture}, page~\pageref{item:modules_ivisit_widgets}), that is, {\tt IVisitSliderWidget},
  {\tt IVisit ListSelectionWidget}, {\tt IVisitTextInputWidget}, and {\tt IVisitCheckboxWidget}. Additional Parameter Widgets contained
  in {\tt ivisit.widgets} (but not displayed in Fig.~\ref{fig:design_gensim}) are now {\tt IVisitDictSliderWidget}, {\tt IVisitRadiobuttonWidget},
  and {\tt IVisitButton Widget}.
\item {\bf Data Widgets} interface with Python variables used for storing simulation outputs (cast as class {\tt DataArray}).
  The subclasses {\tt DataImage} and {\tt DataTxtField}
  displayed in Fig.~\ref{fig:design_gensim} correspond to subclasses also derived from base class {\tt IVisitWidget} in module {\tt ivisit.widgets}
  (see section~\ref{sec:architecture}, page~\pageref{item:modules_ivisit_widgets}), that is, {\tt IVisitImageWidget} and {\tt IVisitTextfieldWidget}.
\item {\bf Comment Widgets} do not interface with Python variables, but just define text fields to explain the system simulated by IVISIT and
  the corresponding parameter and data widgets. The subclass {\tt CommentTxtField} displayed in Fig.~\ref{fig:design_gensim} correspond to
  {\tt IVisitTextCommentWidget} in module {\tt ivisit.widgets}, and is also a subclass of base class {\tt IVisitWidget}.
\end{enumerate}
Finally, class {\bf Simulation} aggregates ParameterWidgets, DataWidgets and CommentWidgets. An IVISIT app can hold many {\tt Simulation} objects, each defining a particular parameter setting vie the class {\tt Parameter} connected to the ParameterWidgets, for example, corresponding to multiple optima or different system regimes. This data is stored in SQL tables (via SQLite
wrapped by supy.sqldatabase).

\begin{figure}[ht]
%\nopagenumber
%\renewcommand{\baselinestretch}{1.0}
\hfill
\begin{center}
\includegraphics[width=\linewidth]{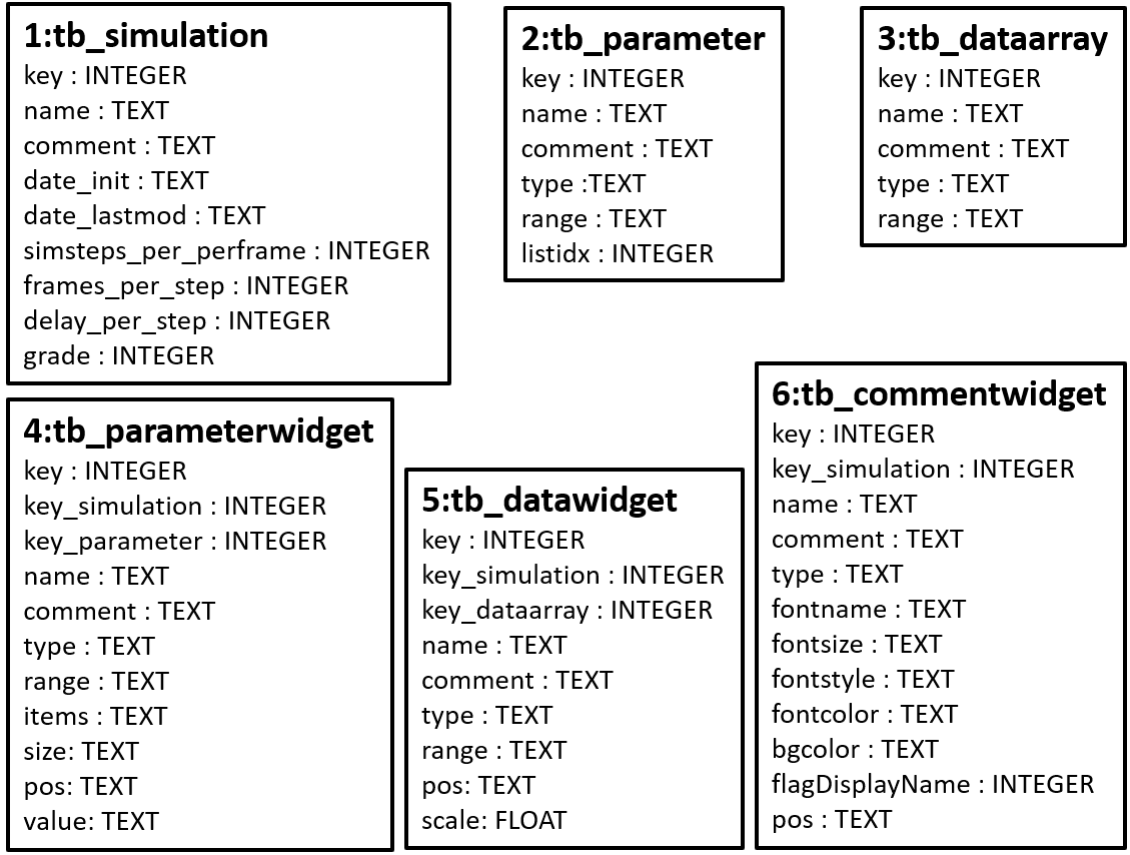}    
\end{center}
\caption{\label{fig:IVISIT_SQL_tables}
  Overview of the SQL tables of IVISIT (current version) as defined in {\tt ivisit.defdb.py}. See text for details.
} 
\end{figure}

Fig.~\ref{fig:IVISIT_SQL_tables} gives an overview of the {\bf SQL database tables} and their attributes as used by the current IVISIT implementation and defined in {\tt ivisit.defdb.py}.
Here the six SQL tables {\tt tb\_simulation}, {\tt tb\_parameter}, {\tt tb\_dataarray}, {\tt tb\_parameterwidget}, {\tt tb\_datawidget}, and {\tt tb\_comment widget} implement the six corresponding main classes {\tt Simulation},  {\tt Parameter},  {\tt DataArray},  {\tt ParameterWidget},  {\tt DataWidget}, and {\tt CommentWidget} of the UML diagram (Fig.~\ref{fig:design_gensim}). In contrast, the various subclasses of {\tt ParameterWidget} and {\tt DataWidget} as displayed in the UML diagram have been implemented in the Python module {\tt ivisit.widgets}. For any IVISIT simulation there is an SQL database file storing the SQL tables. By default, the SQL database file has the same name as the IVISIT simulation, but ending with {\tt $\ast$.db} (see section~\ref{sec:example1_HelloWorld}). 
% Figure fig2_IVISIT_SQL_tables
% created with PowerPoint on HS-Laptop
% --> see: Figur 2
%     in \Users\knoblauch\Desktop\Aktuelles_Draft\KEIM-Projekte\KICAD-CMCENgineers-hermle\Berichte\ivisit_techreport_materials\TechreportIVISIT_Figuren.pptx 
%     Kopie davon liegt auf: ~/hs-albsig/02-research/KICAD/ivisit_techreport/figures/ivisit_techreport_materials
% exportiert als png von PowerPoint; umgewandelt mit img2pdf in pdf

% *******************************************************************************************
% *******************************************************************************************
\section{In-code Python commands to create IVISIT GUI widgets}    \label{app:InCodeCommandsGUIWidgets}   
% *******************************************************************************************
% *******************************************************************************************
IVISIT GUI widgets can in principle be created and modified within the IVISIT application by clicking on the sub-menus of menu {\tt Databases}
and directly inserting, updating, and deleting in the SQL tables (see Fig.~\ref{fig:IVISIT_SQL_tables} in appendix~\ref{app:UML_class_diagrams};
see also {\tt ivisit.widgets.py} in section~\ref{sec:architecture}, page~\pageref{item:modules_ivisit_widgets}).
However, it is much more convenient and recommended to {\bf define IVISIT GUI widgets directly in the Python code} of the IVISIT application
that can be parsed by {\tt ivisit.parser.py} (see examples in section~\ref{sec:examples}). The following gives an overview of the Python in-code commands to define
{\tt IVisitWidget} objects: First a {\bf Simulation Context} (i.e., an entry for SQL table {\tt tb\_simulation}) with an arbitrary name {\tt} should be defined by 
\begin{itemize}\tightitems
\item {\tt \#@IVISIT:SIMULATION \& sim\_name}\\
      Here {\tt sim\_name} is the name of the simulation context.
\end{itemize}
Then {\bf Parameter Widgets} of the types {\tt IVisitSliderWidget}, {\tt IVisitDictSliderWidget}, {\tt IVisit ListSelectionWidget},
{\tt IVisitTextInputWidget}, {\tt IVisitCheckboxWidget}, {\tt IVisitRadiobuttonWidget}, and {\tt IVisitButton Widget} can be defined by:
\begin{itemize}\tightitems
\item {\tt \#@IVISIT:SLIDER  \& name    \& [200,1] \& [0,9,3,1] \& var \& -1 \& int \& 0}    \label{txt:item:SLIDER_directive}\\
  Here {\tt name} is the name of the {\bf SLIDER}; {\tt [200,1]} defines the width of the slider (here 200) and the height (here 1, but currently unused);
  {\tt [0,9,3,1]} defines the slider range min=0, max=9, increment=1 and
  displays 3 equally-spaced tick values on the slider; {\tt var} is the Python parameter variable (defined as attribute of a sub-class of {\tt ivisit.IVisit\_Parameters});
  {\tt -1} corresponds to the parameter list index, if {\tt var} is a list (-1 means that {\tt var} is not a list, but a single value); 
  {\tt int} is the base data type of the Python variable {\tt var} (either {\tt int} or {\tt float}); {\tt 0} is the initial value for the parameter.
\item {\tt \#@IVISIT:DICTSLIDER  \& ParDict  \& [200,20,-1,2,10] \& dict\_par \& 0 }\\
      {\tt \#@IVISIT:DICTSLIDERITEM \& Item1 \& [0, 9,3,1] \& item1 \& int   \& 3}\\
      {\tt \#@IVISIT:DICTSLIDERITEM \& Item2 \& [0,30,4,2] \& item2 \& float \& .5}\\
      The first line produces a {\bf DICTSLIDER} with name {\tt ParDict}
      which is a slider similar as above, but that can manipulate a whole dict of many parameter values, where the parameter value can be selected from an option list.
      {\tt [200,20,-1,2,10]} is the size specification {\tt [SliderWidth, columns,rows,DisplayMode,FontSize]} with the following meaning: \\
      {\tt SliderWidth} is the width of the slider (here 200 pixels);\\
      {\tt Columns} is the width of the text field (in characters), only for DisplayMode=2;\\
      {\tt Rows} is the height of the text field (in characters), only for DisplayMode=2; if -1 then height corresponds to number of dict items;\\
      {\tt DisplayMode} determines the slider display, 0=slider and simple option menu to select the dict items, 1=slider and option menu
      including parameter values, 2=slider, simple option menu, and text field for
      parameter output and input (press ``Set'' to confirm parameter values);\\
      {\tt FontSize} is font size in text field (only for DisplayMode=2).\\
      Lines 2 and 3 each define a {\bf DICTSLIDERITEM}, which specifies the handling of each of the parameters in the dict. The general format is
      {\tt \#@IVISIT:DICTSLIDERITEM \& <ItemName> \& <RangeList> \& <Item> \& <type> \& <value>}. Here {\tt <ItemName>} will be displayed
      in the option menu or text field identifying the parameter to be manipulated by the slider; {\tt <RangeList>=[min,max,nticks,scale]}
      defines the range of the slider similar as for the {\tt SLIDER} widget (see above); {\tt <Item>} is the key of the item in the parameter dict;
      {\tt <type>} is the type of the item (either {\tt int} or {\tt float}); {\tt <value>} is the initial value for parameter.
      There must be at least one {\tt DICTSLIDERITEM} per {\tt DICTSLIDER}.
\item {\tt \#@IVISIT:TEXT\_IN \& name     \& [20,5] \& strvar \& -1 \& InitialText}\\
      Here {\tt name} is the name of the {\bf TEXTIN} text-input-box; {\tt name} defines the name of the text field; {\tt [20,5]} defines the size of the text field as
      20 columns and 5 rows; {\tt strvar} is a Python string variable where text is copied to; -1 indicates that a single string variable is used (if $\ge 0$
      then this number is interpreted as index in a list of string variables); {\tt InitialText} is initial text for the text field.
\item {\tt \#@IVISIT:LISTSEL \& name \&[20,5] \& [A,B,C] \& var \& -1 \& string \& A}\\
  Here {\tt name} is the name of the {\bf LISTSEL} list-selection object; {\tt [20,5]} defines size of 20 columns and 5 rows;
  {\tt var} is a Python variable where the text of the selected option is copied to; -1 indicates that a single string variable is used (if $\ge 0$ 
  then this number is interpreted as index in a list of string variables); {\tt string} is the type of the Python variable (may also be {\tt int} or {\tt float});
  {\tt A} is the initial value. Do not use any quotation marks in these definitions!
\item {\tt \#@IVISIT:CHECKBOX \& name \& [AA,BB,CC,DD] \& strvar \& 0110}\\
  Here {\tt name} is the name of the {\bf CHECKBOX} object; {\tt [AA,BB,CC,DD]} defines a list of possible items that can be selected by checking;
  {\tt strvar} is a Python string variable, where the checkbox selection is copied to, using the format of a '0/1' digit for each item; for example,
  initial value {\tt 0110} means that {\tt BB} and {\tt CC} have been checked and thereby selected (whereas {\tt AA} and {\tt DD} have been unchecked and thereby not selected).
  Do not use any quotation marks in these definitions!
\item {\tt \#@IVISIT:RADIOBUTTON \& name \& [AA,BB,CC,DD] \& strvar \& AA}\\
  Here {\tt name} is the name of the {\bf RADIOBUTTON} object; {\tt [AA,BB,CC,DD]} defines a list of possible items that can be selected by clicking;
  {\tt strvar} is a Python string variable, where the selected item is copied to;
  {\tt AA} is the initial value for the string variable.
  Do not use any quotation marks in these definitions!
\item {\tt \#@IVISIT:BUTTON \& name \& [labeltext,buttontext] \& strvar}\\
  Here {\tt name} is the name of the {\bf BUTTON}, which is otherwise unused; on screen the button
  will be displayed as {\tt labeltext buttontext} (that is {\tt buttontext} is printed on the button widget);
  It is recommended to use {\tt labeltext=<}, so you can move the button on screen by clicking on ``$<$'', whereas clicking on {\tt buttontext} will activate the button;
  if the button is clicked then this will set the Python string variable {\tt strvar} to ``1'' (else to ``0'').
  Do not use any quotation marks in these definitions!
\end{itemize}
{\bf Data Widgets} of the types {\tt IVisitImageWidget} and {\tt IVisitTextfieldWidget} can be defined by:
\begin{itemize}\tightitems
\item {\tt \#@IVISIT:IMAGE     \& name \& 1.0    \& [0,255]  \& img\_var \& int}\\
  Here {\tt name} is the name of the {\bf IMAGE} widget; {\tt 1.0} is the scaling factor; {\tt [0,255]} is the range of the image pixel values;
  {\tt img\_var} is the Python variable for the image matrix (typically a Numpy array) having 2 dimensions (for gray values) and 3 dimensions (for RGB images);
  {\tt int} is the type of the image matrix. After each simulation step, the content of the variable {\tt img\_var} is displayed as image in the IVISIT GUI.
\item {\tt \#@IVISIT:TEXT\_OUT \& name \& [20,5] \& list\_of\_options \& strvar}\\
  Here {\tt name} is the name of the {\bf TEXT\_OUT} widget; {\tt [20,5]} is the size of the text box (20 columns and 5 rows of characters);
  {\tt list\_of\_options} is either {\tt None} or a single string or a list of strings, where possible strings are {\tt just\_left}, {\tt just\_right}, or {\tt just\_center},
  defining the text alignment / justification (default is {\tt just\_center}). 
\end{itemize}
{\bf Comment Widgets} can currently be constructed only within the IVISIT application by clicking on menu item {\tt Databases/Comment\_Widgets}.

\section*{Acknowledgments}   % for the original ideas see A.
This work was partially supported by the Ministerium f\"ur Wirtschaft, Arbeit und Tourismus Baden-W\"urttemberg (VwV Invest BW - Innovation) via
the project KICAD (FKZ BW1\_0092/02),
by the Ministerium f\"ur Wirtschaft, Arbeit und Wohnungsbau Baden-W\"urttemberg (KI-Innovationswettbewerb BW f\"ur Verbundforschungsprojekte) via project KI4Audio (FKZ 36-3400.7/98),
and by the Deutsches Bundesministerium f\"ur Verkehr und digitale Infrastruktur
(Modernit\"atsfonds/mFUND) via the project AI4Infra (FKZ 19F2112C). 

%
% ---- Bibliography ----
%
\setlength{\itemsep}{0cm}
\bibliographystyle{plain}
\bibliography{literatur}

\end{document}